\documentclass[a4paper,fleqn,usenatbib]{mnras}
\usepackage{newtxtext,newtxmath}
\usepackage[T1]{fontenc}
\usepackage{ae,aecompl}
\pdfoutput=1 

\usepackage[latin1]{inputenc}
\usepackage[]{graphicx}	
\usepackage[hyphenbreaks]{breakurl}
\usepackage{amsmath}	
\usepackage{amssymb}	

\usepackage[lofdepth,lotdepth]{subfig}
\usepackage{lscape}
\usepackage{placeins}
\usepackage{xspace}
\usepackage{color,xcolor,ucs}

\DeclareRobustCommand{\Chi}{{\mathpalette\irchi\relax}}
\newcommand{\irchi}[2]{\raisebox{\depth}{$#1\chi$}}

\newcommand{\refeq}[1]{Eq. (\ref{#1})}
\newcommand{\refeql}[1]{(see Eq. \ref{#1})}
\newcommand{\refeqs}[3]{Eqs. (\ref{#1}),  (\ref{#2}) and (\ref{#3})}
\newcommand{\refsec}[1]{Section \ref{#1}}
\newcommand{\refsecl}[1]{(see Sec. \ref{#1})}
\newcommand{\refim}[1]{Fig. \ref{#1}}
\newcommand{\refapp}[1]{Appendix \ref{#1}}
\DeclareMathOperator\arctanh{arctanh}
\DeclareMathOperator\sinc{sinc}
\newcommand{\avg}[1]{\left\langle #1 \right\rangle}
\newcommand{\abs}[1]{\left| #1 \right|}
\newcommand{\order}[1]{\mathcal{O}\left(#1\right)}
\newcommand{\Mpc}{$h^{-1}$Mpc\xspace}
\newcommand{\bx}{\boldsymbol{x}}
\newcommand{\Msun}{$h^{-1}M_\odot$\xspace}
\newcommand{\ie}{i.e.\xspace}

\newcommand{\varsc}{\mathcal{R}}
\newcommand{\vart}{\tau}
\newcommand{\varvel}{\Delta_{vel}}
\newcommand{\grad}[1]{\bold{\nabla}#1}

\makeatletter
\let\@fnsymbol\@arabic
\makeatother

\title[Reconstructing matter profiles of CoSpheres]{Reconstructing matter profiles of spherically compensated cosmic regions in $\Lambda$CDM cosmology}
\author[Paul de Fromont $\&$ Jean-Michel Alimi]{Paul de Fromont\textsuperscript{\thanks{email:paul.de-fromont@obspm.fr}}Jean-Michel Alimi\textsuperscript{\thanks{email:jean-michel.alimi@obspm.fr}}\\
LUTH, Observatoire de Paris, PSL Research University, CNRS, Université Paris Diderot, Sorbonne Paris Cité\\
5 place Jules Janssen, 92195 Meudon}

\date{Accepted XXX. Received YYY; in original form ZZZ}

\pubyear{2017}

\begin{document}
\label{firstpage}
\pagerange{\pageref{firstpage}--\pageref{lastpage}}
\maketitle

\begin{abstract} 
The absence of a physically motivated model for large scale profiles of cosmic voids limits our ability to extract valuable cosmological information from their study. In this paper, we address this problem by introducing the spherically compensated cosmic regions, named CoSpheres. Such cosmic regions are identified around local extrema in the density field and admit a unique compensation radius $R_1$ where the internal spherical mass is exactly compensated. Their origin is studied by extending the standard peak model and implementing the compensation condition. Since the compensation radius evolves as the Universe itself, $R_1(t)\propto a(t)$, CoSpheres behave as bubble Universes with fixed comoving volume. Using the spherical collapse model, we reconstruct their profiles with a very high accuracy until $z=0$ in N-body simulations. CoSpheres are symmetrically defined and reconstructed for both central maximum (seeding haloes and galaxies) and minimum (identified with cosmic voids). 
We show that the full non linear dynamics can be solved analytically around this particular compensation radius, providing useful predictions for cosmology. This formalism highlights original correlations between local extremum and their large scale cosmic environment. The statistical properties of these spherically compensated cosmic regions and the possibilities to constrain efficiently both cosmology and gravity will be investigated in companion papers.
\end{abstract}

\begin{keywords}  
cosmology: theory; large-scale structure of Universe; N-body Simulations; Cosmic Voids; dark energy
\end{keywords}  

\section*{Introduction}

One of the main purpose of modern cosmology is to understand the nature of Dark Energy (DE), driving the cosmic acceleration \citep{Riess1998, Perlmutter1999, Silvestri2009, Caldwell2009}. It is not only difficult to build consistent models to understand this acceleration but rather to find binding limits to discriminates between them.

Large Scale Structures (LSS) offer a large panel of probes for cosmology and the nature of gravity itself. They carry informations on both primordial Universe and gravity through the cosmological evolution. These last years, cosmic structure formation have been specially studied in the frame of cosmic voids formation and statistics \citep{Park2007, Lavaux2012, Pan2012, Hamaus2015, Cai2015, Achitouv2016, Hamaus2016,Achitouv2017}. Although cosmic void dynamics is far from being linear, these regions are safe from the highly non linear physics occurring during haloes or galaxy formation. Moreover, cosmic voids are expected to be more sensitive to the nature of DE since their local $\Omega_{DE}$ is higher than in the average Universe \citep{Sheth2004, Colberg2005, Vandeweygaert2011}. Voids have been studied through the Alcock-Paczynski test \citep{Sutter2014a, Mao2016}, their ellipticity \citep{Park2007, Lavaux2010} and their abundance or shape \citep{Cai2015, Achitouv2015}. However, all these studies suffer from the lack of a fully consistent - and physically motivated - model describing both the origin and the dynamical generation of such cosmic regions. \citet{Hamaus2014} introduced an effective parametrization of density profiles using numerical simulation. Despite not being deduced from first principles, it can provide physical insights. For example, \citet{Hamaus2016} used it to model the isotropic shape of the void-density profile and have been able to isolate the sensibility to cosmological parameters through anisotropic redshift-space and Alcock-Paczynski distortions.

\newpage
In this paper and through the following ones \citep{paper2, paper3, paper4} we present a physically motivated model studying both the primordial origin and the dynamical evolution of such cosmic regions. More precisely, we generalize cosmic void study by introducing the spherically compensated cosmic regions, named thereafter CoSpheres\footnote{For \textbf{Co}mpensated \textbf{Spher}ical regions}. These structures are defined as the large scale cosmic environment surrounding local extrema of the density field. When defined around central under densities (local minimum), these regions can be identified to cosmic voids. Interestingly, these regions can also be defined around central over dense maxima, defining the symmetric of standard voids.

In average\footnote{we discuss this term more precisely in this paper}, the large scale environment around maxima (respectively minima) in the density field can be separated in two distinct domains: an internal over (respectively under) dense core surrounded by a large under (respectively over) dense compensation belt. Note that even if the density field around local extrema is far from being spherical, one can always define a spherical profile by averaging over angles. However, despite being intuitive, the density contrast $\delta(\bx)=\rho(\bx)/\bar{\rho}_m-1$ has no dynamical interpretation. Indeed, in the spherical frame, the local gravitational dynamics is driven by the integrated density contrast, or equivalently the mass contrast
\begin{equation}
\label{Delta0}
\Delta(r)=\frac{3}{r^3}\int_0^ru^2\delta(u)du = \frac{m(r)}{4\pi/3\bar{\rho}_mr^3}-1
\end{equation}
Like for density, the large scale environment of local extrema can be splitted in an over-massive (respectively under-massive for central minima) core surrounded by an under massive (respectively over massive) area. The transition radius between these under/over massive regions defines the \textit{compensation radius}, noted $R_1$. This radius can be uniquely define for each central extrema\footnote{more generally, for any random position in the density field}. In a naive spherical description, this radius separates the collapsing over massive region from the expanding under massive one. The existence of such scale is fundamentally insured by the Bianchi identities \citep{Hehl1986} which impose the mass conservation. Moreover, the compensation radius $R_1$ follows a remarkable evolution. Indeed, since $R_1$ encloses a sphere whose averaged density equals the background density, it evolves as the scale factor itself, \ie $R_1(t)\propto a(t)$. CoSpheres thus behave as bubble universes with a fixed comoving size. \citet{Hamaus2014a} introduced a similar concept of a compensation radius for voids and its use as a static cosmological ruler that follows the background expansion. However, our definition of the compensation radius differs since it is defined uniquely for each maximum \refeql{R1}. We also stress that, on the Hubble size, there should not "over-compensated" or "under-compensated" voids as a consequence of the mass conservation.

The large scale structures are originally generated by the stochastic fluctuations of the density field in the primordial Universe. Their statistical properties, including average shape and probability distribution can be computed within the Gaussian Random Field (GRF) formalism. However, it is necessary to implement the compensation constraint (the existence of a finite compensation radius $R_1$) and thus to extend the results of \citet{BBKS}. As we show in this paper, the non linear evolution of such regions is very well described by using the spherical collapse model while neither Zel'dovich nor Eulerian linear dynamics is accurate enough.

We discuss the linear scaling of the density profiles of such regions in both primordial and evolved Universe. It turns out that these large scale profiles do not scale linearly on $R_1$, neither on shape nor amplitude. This property emerges from the fact that on scales considered here (from $r\sim 5$ to $r\geq 100$ \Mpc), the linear matter power spectrum is far from being scale invariant. Moreover, the non linear gravitational evolution of these profiles would have broken any primordial linear scaling.

The paper is organized as follow. In \refsec{sec:cosphere} we introduce the N-body simulations on which is based our study; the DEUS simulations \refsecl{sec:DEUS}. After defining precisely CoSpheres, we study these regions in the numerical simulations for various redshift and sizes and for both central over and under densities. This leads us to discuss the stacking method used to reconstruct the corresponding average profiles.

In \refsec{sec:initial} we study the shape of these regions in a Gaussian primordial field. We present an extension of the usual peak formalism of BBKS \citep{BBKS}. While BBKS formalism focuses on the local properties of the field around the peak (note that for us, a peak is an extremum and can be a minimum or a maximum), we extend this model to take into account its cosmic environment on large scale. We show that this environment can be fully qualified by the compensation scale $R_1$ and the compensation density $\delta_1=\delta(R_1)$.

In \refsec{sec:dynamic} we  study the dynamical evolution of CoSpheres. We show that the Lagrangian Spherical Collapse (SC) model \citep{Padmanabhan1993,Peacock1998} is able to reproduce precisely the evolution of such regions from small scales (typically $r\sim 5$ \Mpc) to much larger scales where the dynamics becomes almost linear. However, we explicitly show that neither the Eulerian linear theory nor the Zel'dovich approximation are able to describe their evolution with a sufficient precision. Finally, we show that we are able to reproduce the full matter field surrounding both maxima (build around haloes) and minima (identified to cosmic voids) at $z=0$ in numerical simulations.

\section{CoSpheres in the Numerical Simulations}
\label{sec:cosphere}

\subsection{N-body DEUS simulations}
\label{sec:DEUS}

In this work we use the numerical simulations from the ``Dark Energy Universe Simulation '' (DEUS) project. These simulations are publicly available through the  ``Dark Energy Universe Virtual Observatory ''  DEUVO Database\footnote{http://www.deus-consortium.org/deus-data/}. They  consist of N-body simulations of Dark Matter (DM) for realistic dark energy models. For more details we refer the interested reader to dedicated sections in \citet{Alimi2010, Rasera2010, Courtin2010, Alimi2012, Reverdy2015}. These simulations have been realized with an optimized version \citep{Alimi2012, Reverdy2015} of the adaptive mesh refinement code RAMSES based on a multigrid Poisson solver \citep{Teyssier2002,Guillet2011} for Gaussian initial conditions generated using the Zel'dovich approximation with MPGRAFIC code \citep{Prunet2008} and input linear power spectrum from CAMB \citep{Lewis2000}. 

In this paper we focus only on the flat $\Lambda$CDM model with cosmological parameters calibrated against measurements of WMAP 5-year data \citep{Komatsu2009} and luminosity distances to Supernova Type Ia from the UNION dataset \citep{Kowalski2008}. The reduced Hubble constant is set to $h=0.72$ and the cosmological parameters are $\Omega_{DE}=0.74$, $\Omega_b=0.044$, $n_s=0.963$ and $\sigma_8=0.79$. In this paper, we used mainly two different simulations whose properties are summarized in Table \ref{tab:simus}.

\begin{table*}
\begin{tabular}{cccc}
\hline 
  & $L=648,$ $n=1024^3$ & $\bold{L=2592,}$ $\bold{n=2048^3}$ & $L=5184,$ $n=2048^3$ \\ 
\hline
$m_p$ in \Msun & $\sim 1.8 \times 10^{10}$ & $\bold{\sim 1.5\times 10^{11}}$ & $\sim 1.2\times 10^{12}$\\
$M_h$ in \Msun & $\sim 4.0\times 10^{12}$ & $\bold{\sim 3.0\times 10^{13}}$ & $\sim 2.5\times 10^{14}$ \\ 
\hline 
\end{tabular} 
\caption{Simulations used in this paper. $m_p$ is the mass of each particle while $M_h$ corresponds to the average mass of the DM haloes selected for stacking. Each simulation is defined by its box size $L$ in \Mpc and the number of DM particles $n$. The simulation in bold is the reference simulation.}
\label{tab:simus}
\end{table*}

If not specified, the numerical simulation used is the reference one defined with $L_{box}=2592$ $h^{-1}Mpc$ and $n_{part}=2048^3$ to get both a large volume and a good mass resolution (see Table \ref{tab:simus}).

\subsection{Defining CoSpheres}

\subsubsection{Method and definition}
\label{method}

We construct CoSpheres in numerical simulation from the position of local extrema in the density field. For central maxima and for $z=0$, we decide to identify these positions with the center of mass of DM haloes\footnote{detected by using a Friend-Of-Friend algorithm with a linking length $b=0.2$} . Such procedure is motivated by the possibility to extend it for observational data where DM haloes could be identified with galaxy, galaxy group or galaxy cluster. 

In the symmetric case of a central minima, the position is computed as local minima in the density field smoothed with a Gaussian kernel. In the reference simulation, the physical size of the original coarse grid cell is $L_{grid}=1.26$ \Mpc before any refinement and we choose a smoothing scale of $5$ \Mpc. The comparison with analytical predictions requires to smooth the matter power spectrum on the same scale.

\begin{figure} 	
	\centering \includegraphics[width=1.0\linewidth]{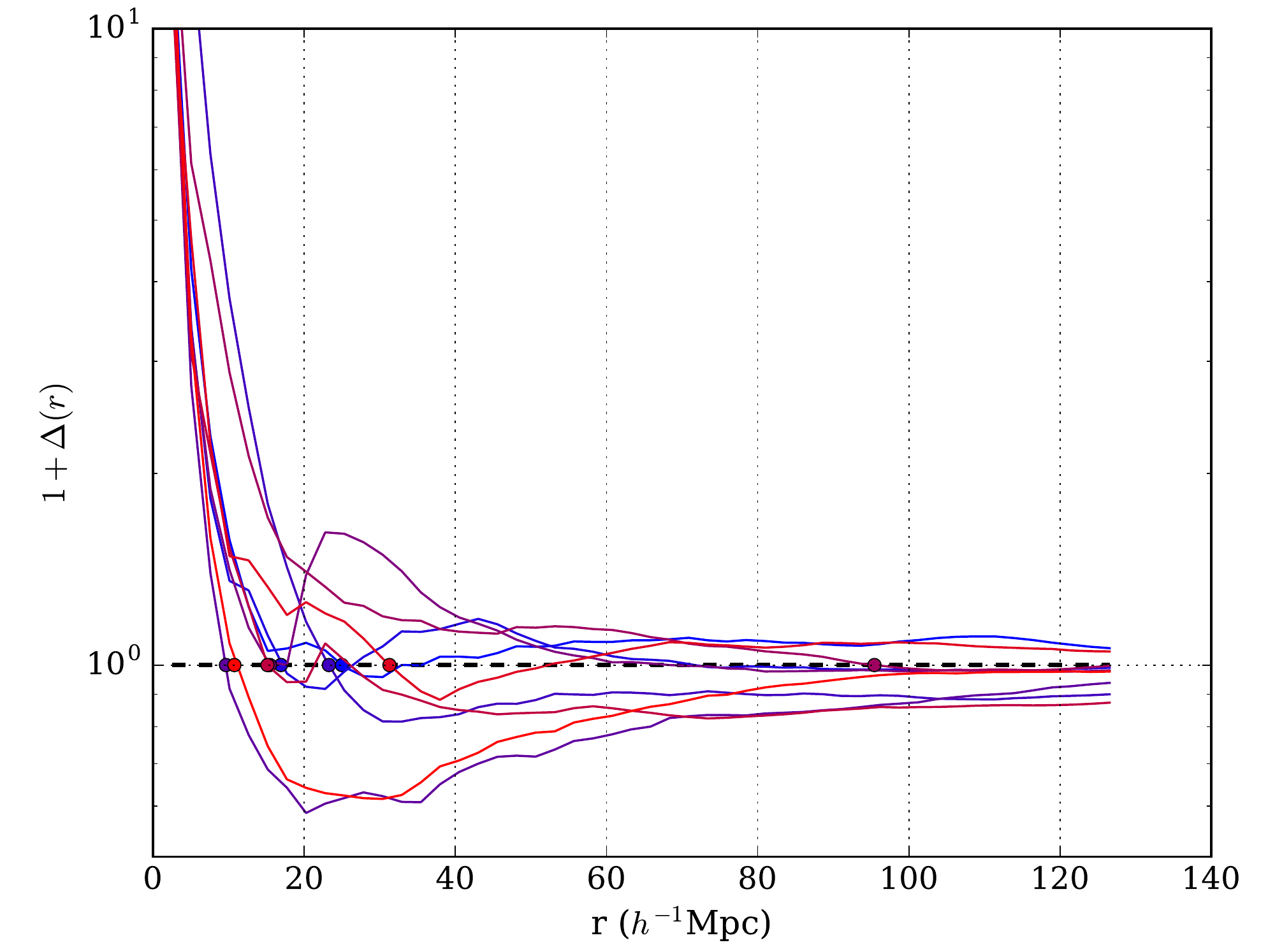}
	\caption{Mass contrast profile built around various haloes of mass $M_h\sim 3\times 10^{13}$ \Msun from the reference simulation at $z=0$ (see Table \ref{tab:simus}). For each profile we show the compensation radius $R_1$ defined by \refeq{R1} and marked with a colored dot. The majority of profiles are compensated on scale between $10$ and $30$\Mpc. We also show a profile with a very large compensation radius $R_1\sim 100$. For central minima, \ie cosmic voids we obtain similar profiles with finite compensation radii, but in this case, both $\delta(r)$ and $\Delta(r)$ are bounded to $-1$.}
	\label{fig:profile_D_R1}
\end{figure}

Around each extremum of the density field, we compute the concentric mass by counting the number of particles in the sphere of radius $r$, thus imposing the spherical symmetry
\begin{equation}
m(r)=\sum_i m_p \Theta\left[r-\abs{\bx_0-\bx_i}\right]
\end{equation}
where $m_p$ is the mass of each individual particle, $\bx_0$ the position of the extremum and $\bx_i$ the position of the $i^{\text{th}}$ particle. $\Theta(x)$ is the standard Heaviside function such as $\Theta(x)=1$ for $x>0$ and $0$ elsewhere. From this mass profile we define the mass contrast profile $\Delta(r)$ defined in \refeq{Delta0}. Note that the density contrast is linked to the mass contrast by
\begin{equation}
\label{structural_relation}
\delta(r)=\frac{1}{3}\frac{\partial \Delta}{\partial \log r} + \Delta(r)
\end{equation}
We now focus on the compensation property. A volume $V$ is said to be compensated if it satisfies the condition
\begin{equation}
\int_V \rho_m(\bold{x})d^3\bold{x} = \bar{\rho}_mV
\end{equation}
The spherical symmetry imposes that the field is compensated in a sphere of radius $R_1$ if it satisfies
\begin{equation}
\label{R1}
m(R_1)=\frac{4\pi}{3}\bar{\rho}_mR_1^3\quad \Leftrightarrow \quad \Delta(R_1)=0
\end{equation}
this last equation defines the compensation scale $R_1$ as the \textit{first} radius satisfying $\Delta(R_1)=0$. We stress that this scale is much larger than the typical scale associated to haloes such as the virialisation radius $R_{vir}$ or $r_{200}$ such as $\rho(r_{200})=200\times\bar{\rho}_m$ \citep{Ricotti2007}. It is important to note that the compensation radius is defined uniquely for each structure despite the fact that the mass contrast may vanish at other radii $R_i>R_1$. For each central extremum, there is a - possibly infinite - number of radii satisfying $\Delta(R_i)=0$ ; the compensation radius is defined as the smallest one. Moreover, since these regions must be compensated on the size of the Universe (no mass excess), we have also $\lim_{r\to\infty}r^3\Delta(r)\to 0$. In \refim{fig:profile_D_R1} we show various mass contrast profiles $1+\Delta(r)$ centred on DM haloes at $z=0$. For each profile we identify $R_1$ where the mass is exactly balanced.

We note that compensation radii are always much smaller than the size of the computing box. In the reference simulation, $70\%$ of profiles are compensated on $R_1\leq 50$ \Mpc whereas less than $7\%$ of the profiles have $R_1\geq 100$ \Mpc. It means that the compensation radius could be also measurable on observational data with a sufficiently large volume survey. Moreover $R_1$ is roughly of the same order of the effective size $R_{eff}$ used in the study of cosmic voids \citep{Platen2007, Neyrinck2008}. 

\subsubsection{Average profile at $z=0$ and stacking procedure}
\label{sec:construction}

Every compensated region detected in numerical simulation is characterized by two distinct properties. One concerning the central extremum fully described by its height (\ie the mass of the halo for a maxima and the central $\delta(\bx_0)$ for minima). The second concerning its cosmic environment, characterized by $R_1$. Numerical simulation provides an ensemble of profiles with various heights and radii. 

\begin{figure}
	\includegraphics[width=1.0\linewidth]{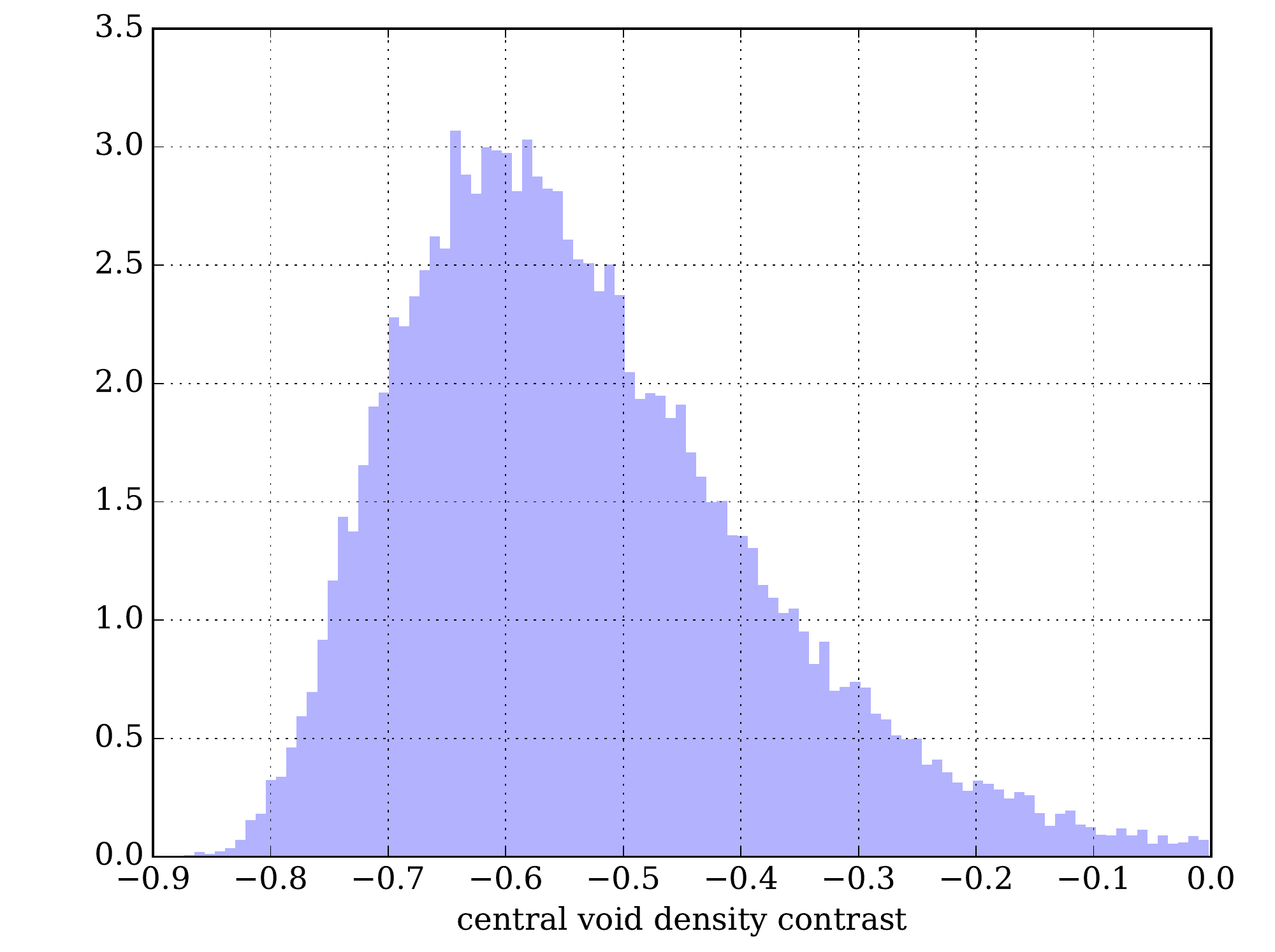}
	\caption{Distribution of the central density contrast $\delta(\bx_0)$ of $50000$ voids at $z=0$ in the reference simulation.}
	\label{fig:central_void_density}
\end{figure}

Due to the stochastic nature of the density field, the only physically relevant elements are obtained by computing average quantities and their dispersion. This leads to defined average spherical profiles. All along this paper, average profiles and their corresponding dispersion are built from at least $3000$ single profiles. This number insures a fair statistical estimation. These profiles are built by stacking together CoSpheres with the same height\footnote{the same halo mass for maxima} and the same compensation radius $R_1\pm dR_1$ where the radial width is $dR_1 = 1.25$ \Mpc. This radial bin is kept constant for the whole paper. For central minima detected in the smoothed density field \refsecl{method}, we stack together profiles with the same $R_1\pm dR_1$ without density criteria except $\delta(\bx_0)<0$. In \refim{fig:central_void_density} we plot the distribution of their central density contrasts (whatever $R_1$). We observe that more than $99\%$ of central contrasts are lower -0.1, beyond Poissonian fluctuations. The resulting profiles are thus averaged over all possible realization of the field with a fixed compensation radius. 

\begin{figure*}
	\captionsetup[subfigure]{width=0.48\linewidth}
	\begin{center}
		\subfloat[Stacked average profile measured around haloes of mass $M_h\sim 3.0\times 10^{13}$ \Mpc at $z=0$ in the reference simulation. We clearly identify the central over-dense core until $r_1$ (red dot) surrounded by the \textit{compensation belt} from $r=r_1$. The same occurs for the mass contrast profile (in blue), \ie an over massive core for $r\leq R_1$ (blue square) enclosed in a large under massive region.]{
			\includegraphics[width=0.5\textwidth]{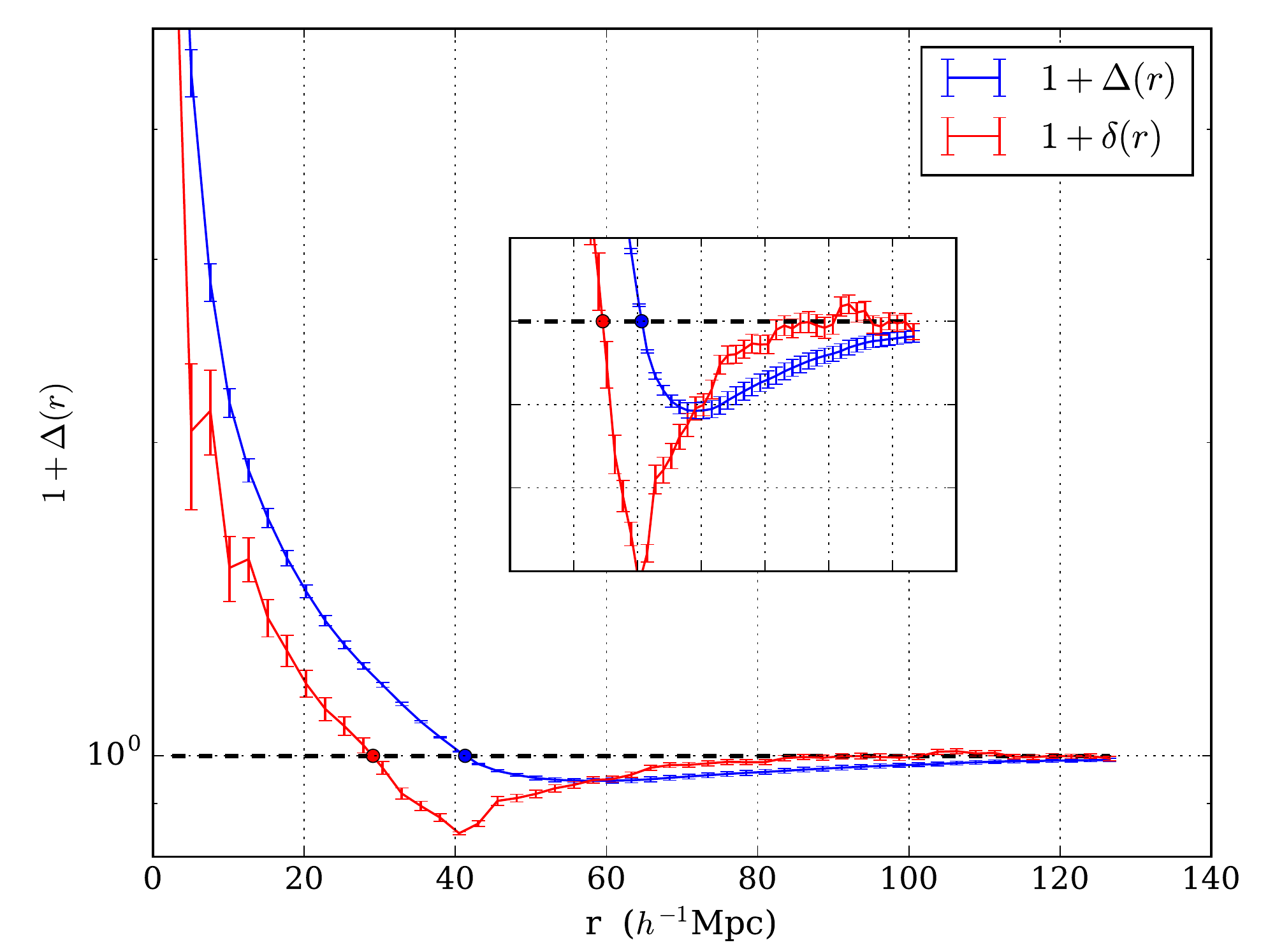}
			\label{fig:profile_D_d_today}}
		\subfloat[Same as the left panel for central minima. Now the interior region $r<R_1$ is under massive while the exterior region is over massive. The compensation radius has been chosen with the same value than in the left panel.]{
			\includegraphics[width=0.5\textwidth]{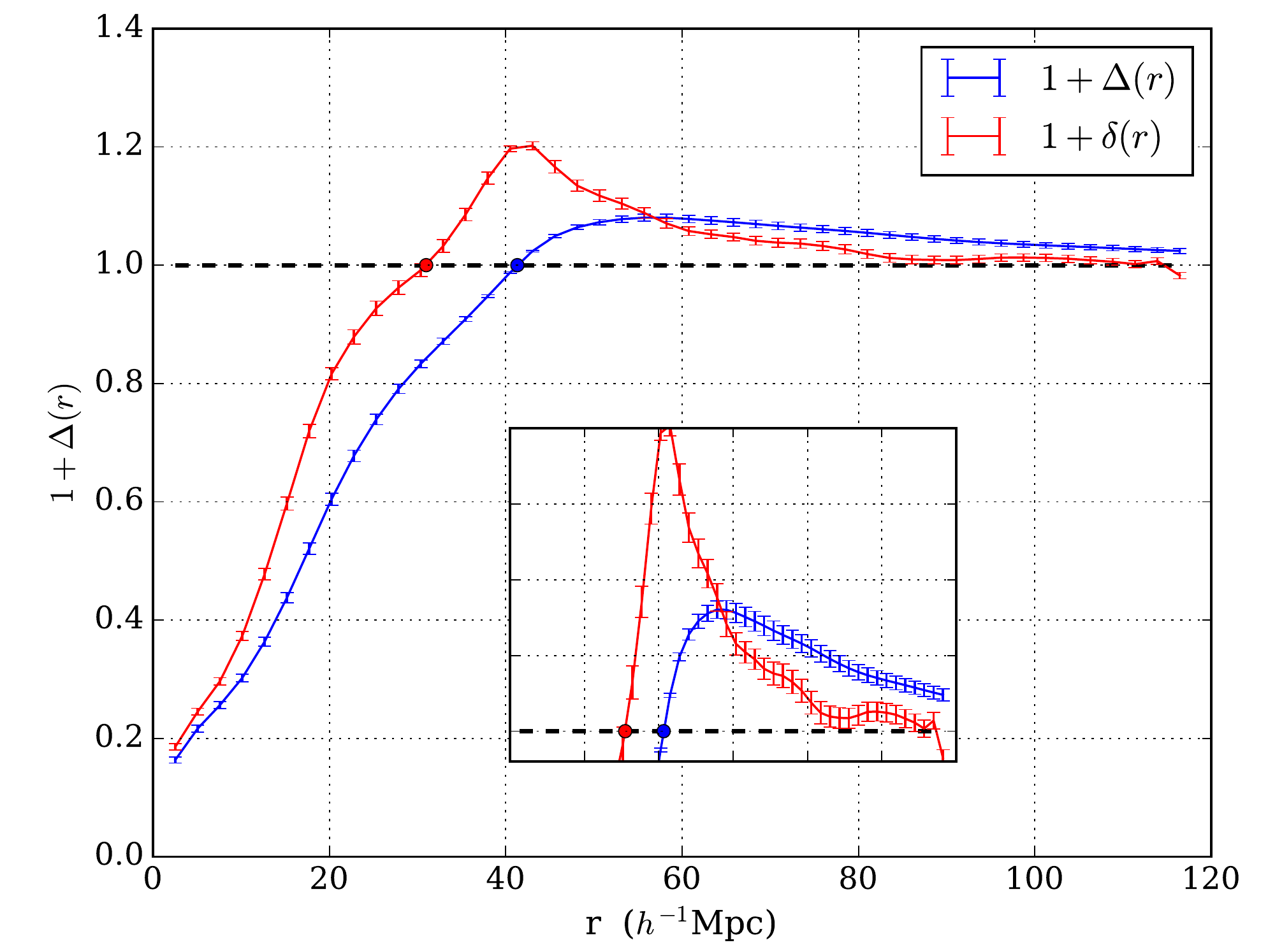}
			\label{fig:profile_D_d_today_void}}
		\caption{Average mass and density contrasts. The blue line represents the mass contrast $\Delta(r)$ while the red line represents the density contrast $\delta(r)$. The density radius (red dot at $r_1\sim 27$ \Mpc) and the mass radius (blue square at $r=40$ \Mpc) can be clearly identified. On both panels, we plot a zoom of profiles around the compensation radius.}
		\label{fig:averaged_profile}
	\end{center}
\end{figure*}

On \refim{fig:averaged_profile} we show average profiles in the reference simulation at $z=0$ from both halo and void with a given compensation radius $R_1=40$\Mpc. In both cases we show the various radii
\begin{itemize}
	\item the density radius $r_1$ such that $\delta(r_1)=0$ (on this figure we have $r_1\simeq 30$ \Mpc). It separates the over and under dense areas.
	\item the compensation radius $R_1$. Note that by construction it satisfies $R_1\geq r_1$ since it encloses an over and a under dense shell (such that they compensated each other).
\end{itemize}
Error bars are computed as the standard error on the mean, \ie $\sigma/\sqrt{n}$ where $\sigma$ is the dispersion and $n$ the number of profiles considered.

\begin{figure*}
	\captionsetup[subfigure]{width=0.48\linewidth}
	\begin{center}
		\subfloat[Stacked average profiles around haloes with a mass $M_h\sim 3.0\times 10^{13}$ \Mpc at $z=0$ in the reference simulation.]{
			\includegraphics[width=0.5\textwidth]{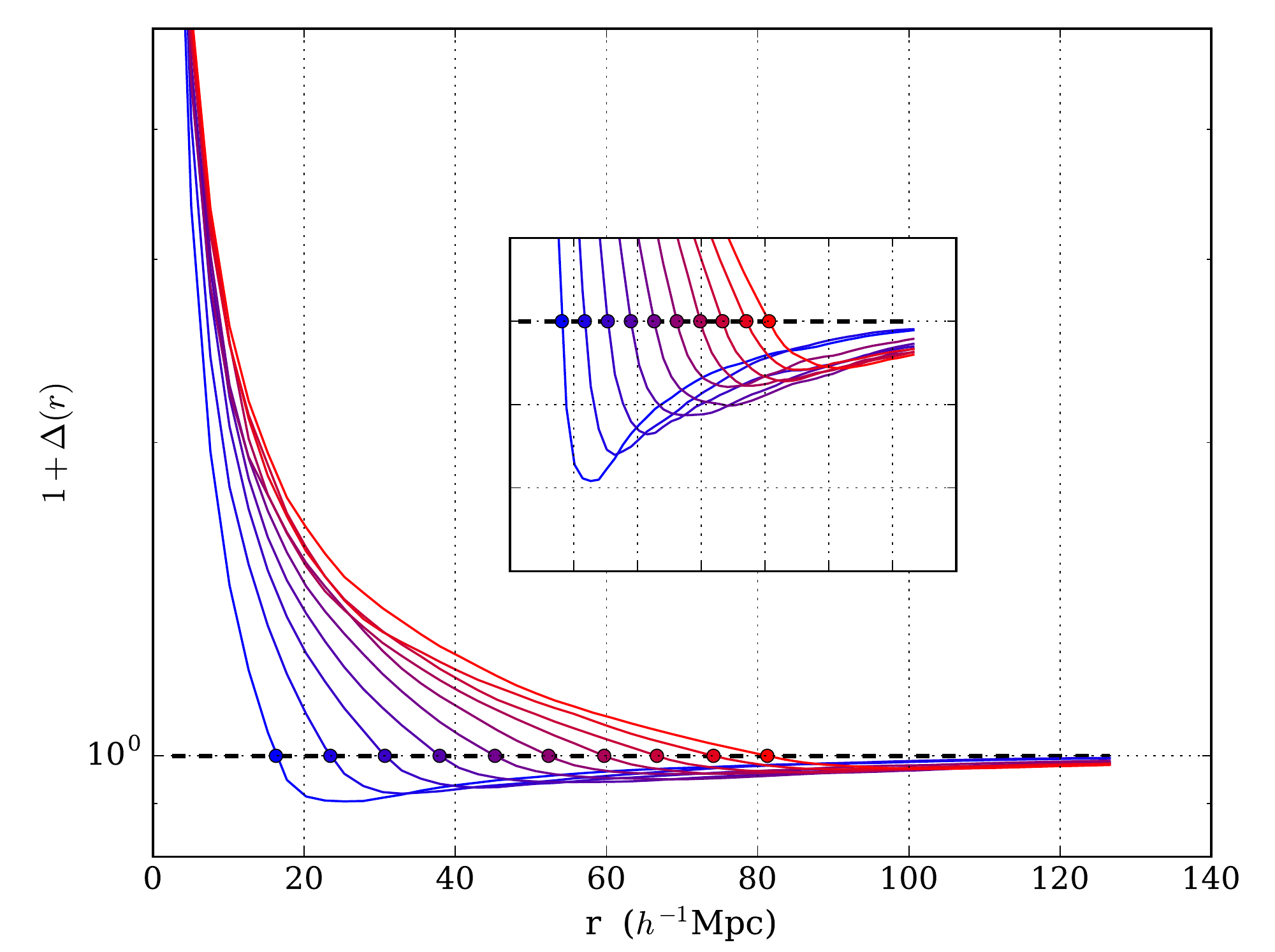}
			\label{fig:profile_D_today}}
		\subfloat[Same as in left panel for central under dense regions, \ie cosmic voids]{
			\includegraphics[width=0.5\textwidth]{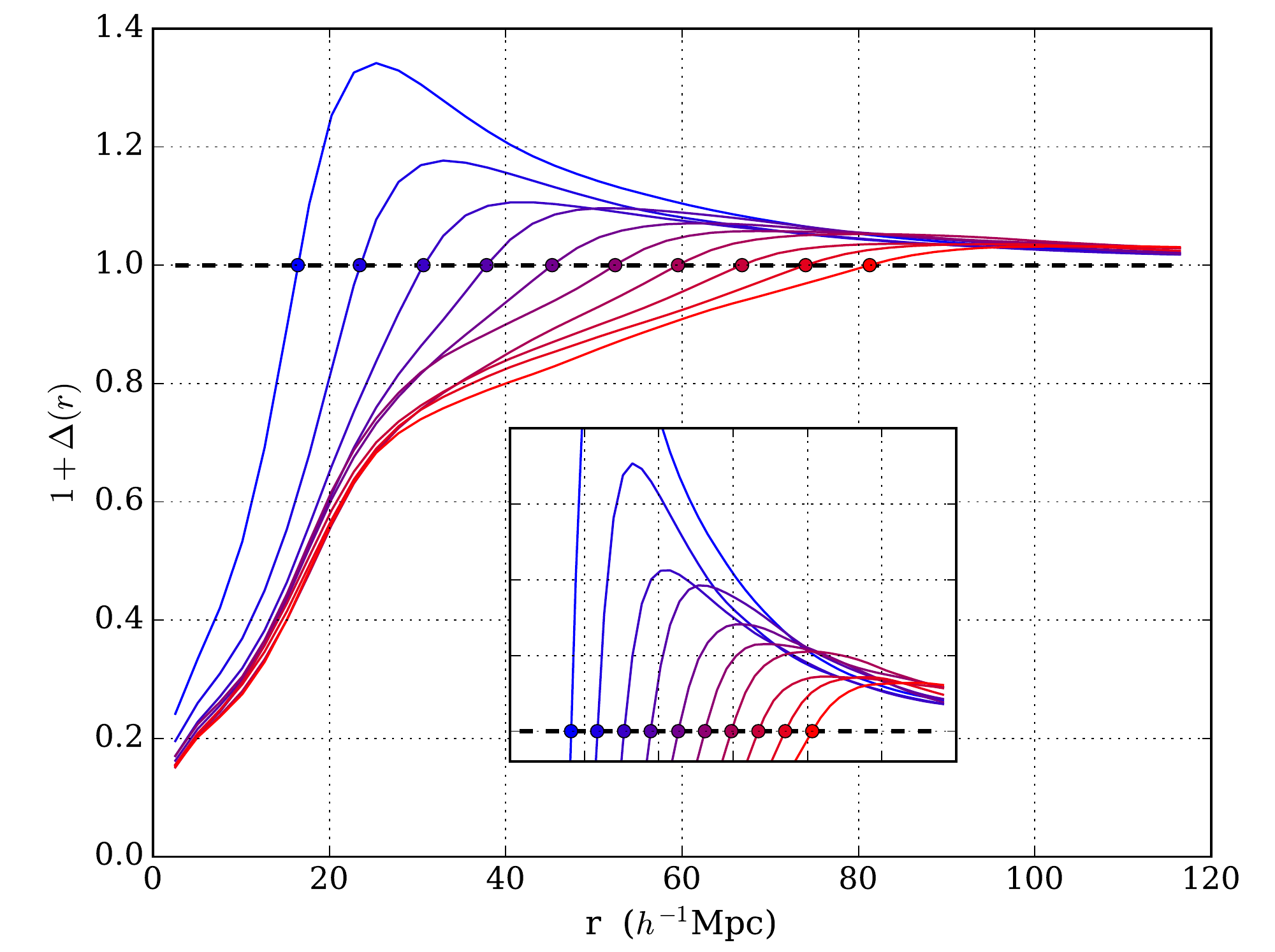}
			\label{fig:profile_D_void_today}}
		\caption{Radial average mass contrast profiles at $z=0$. Each curve corresponds to a fixed compensation scale from $15$ to $80$ \Mpc. We do not show the error bars on this figure since they are almost indistinguishable from the curve itself. In both cases we note that small $R_1$ are associated to strongly contrasted regions.}
		\label{fig:profile_D_today_set}
	\end{center}
\end{figure*}

On \refim{fig:profile_D_today_set}, we plot the stacked average profiles for various compensation radii $R_1$ with the same central extrema. Varying $R_1$ probes the same peak in various cosmic environments.

\begin{figure}
	\includegraphics[width=1.0\linewidth]{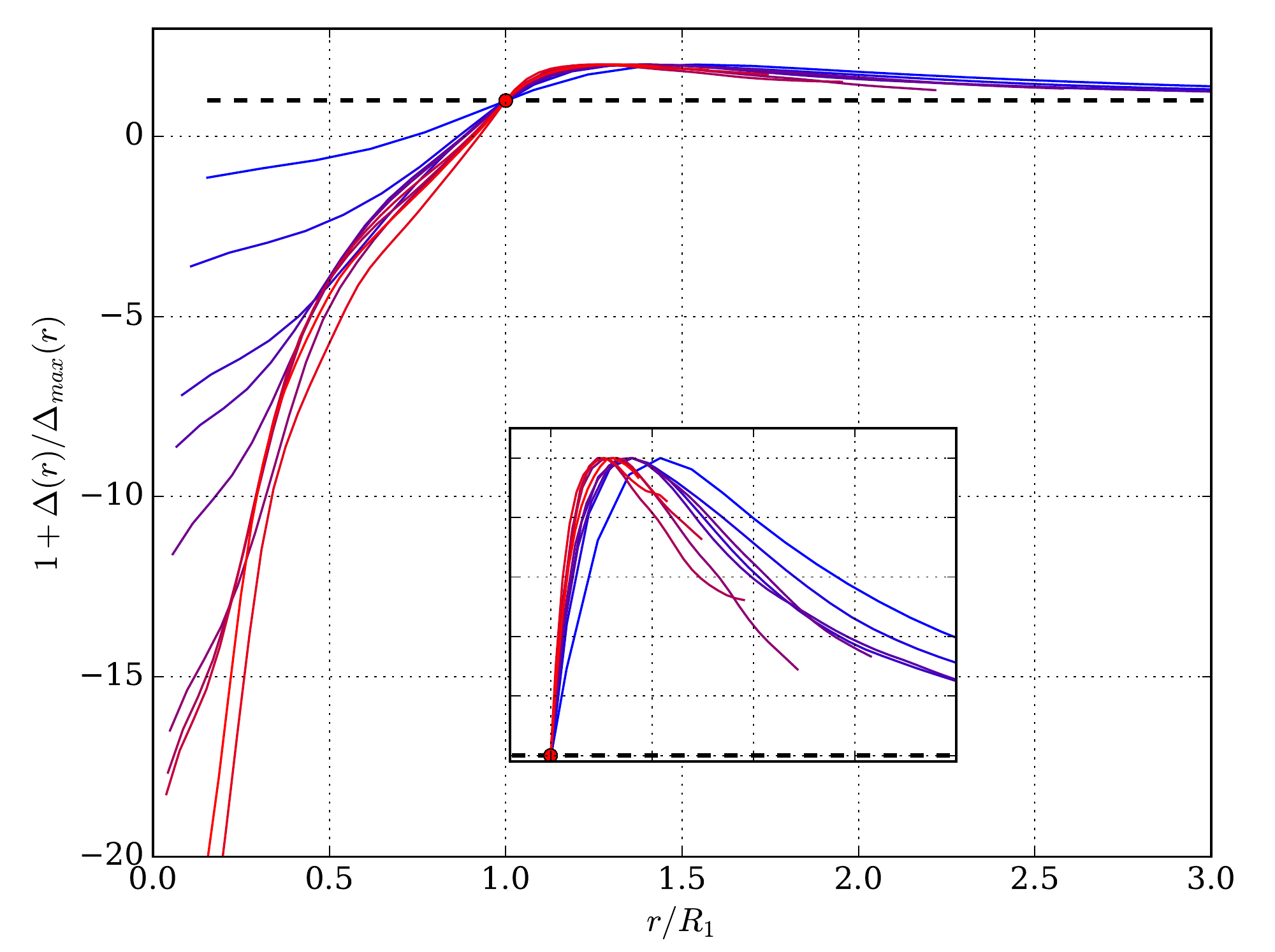}
	\caption{Normalized mass contrast profile $\Delta(r/R_1)/\Delta_{max}$ (where $\Delta_{max}$ is the maximum of each profile) for cosmic voids at $z=0$. We plot the same regions than in \refim{fig:profile_D_void_today}, with $R_1$ from $15$ to $80$ \Mpc (red curve). The clear separation between various profiles shows the explicit non linear dependence of both shape and amplitude in term of $R_1$ and rules out a simple linear scaling of such radial profiles.}
	\label{fig:profile_universality}
\end{figure}

Using these profiles we can study the simple linear scaling assumption. For the mass contrast for example, there could exist $\Delta_{univ}(r)$ such that for any $R_1$ we would have $\Delta(r,R_1)=\alpha \Delta_{univ}(\beta r)$. On \refim{fig:profile_universality}, we plot the rescaled profiles $\Delta(r/R_1)/\Delta_{max}$ where $\Delta_{max}$ is the maximum of the mass contrast. this figure does not indicates any simple linear scaling. Despite being normalized to the same maximal amplitude, the profiles are clearly separated on small scales (for $r\leq R_1$) but also on larger scales. Furthermore, the position of the maximum changes while varying $R_1$, indicating that $R_{max}\not\propto R_1$. This show that it is necessary to study the shape of these regions for various compensation radii.

\begin{figure}
	\includegraphics[width=1.0\linewidth]{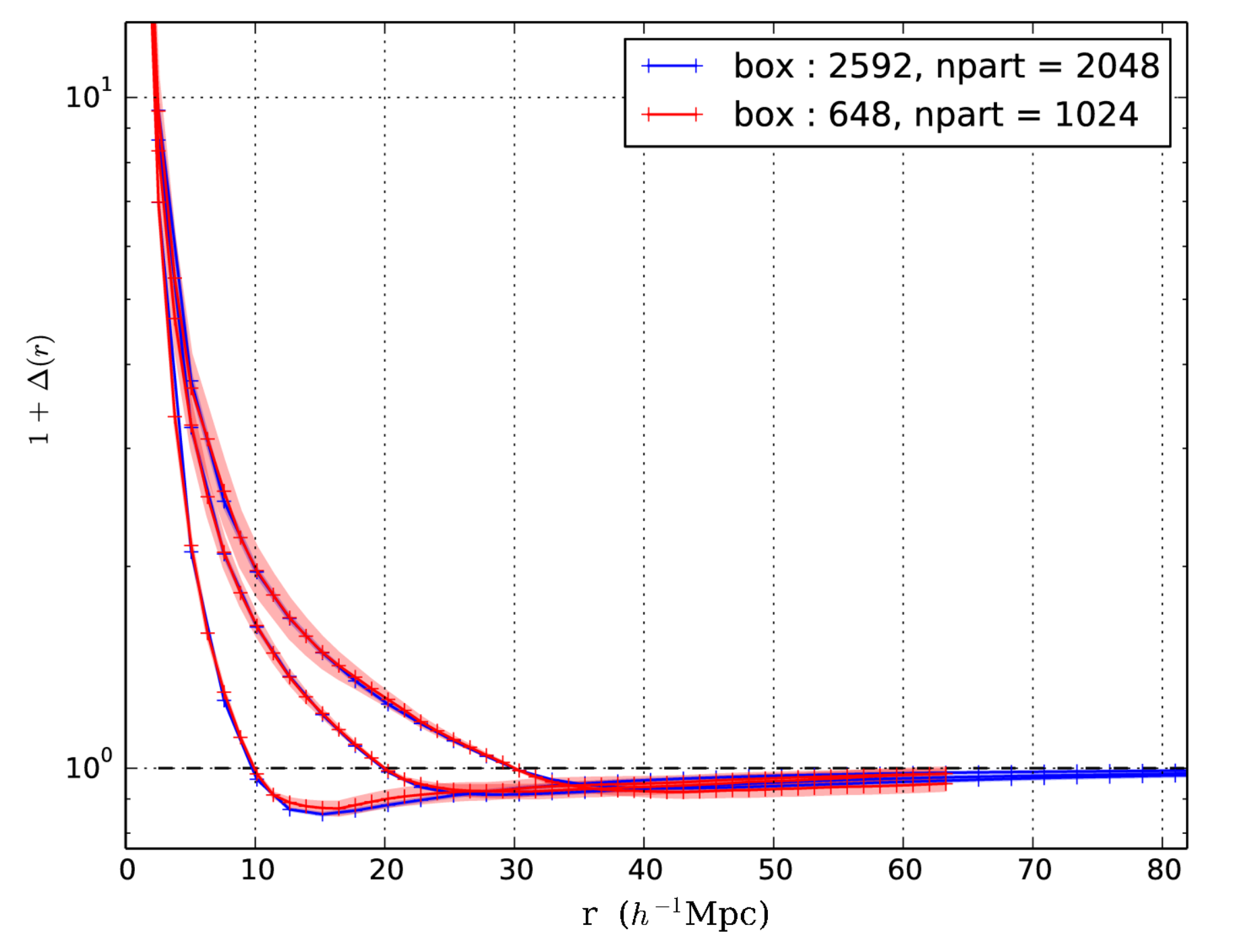}
	\caption{Mass contrast profiles $\Delta(r) + 1$ for three different $R_1$ and for two different $\Lambda$CDM simulations using haloes with the same mass $M_h\sim 1.5\times 10^{13}$ \Msun at $z=0$. The shaded regions show the very low dispersion due to the respective stacking in each simulation. Whatever the simulation, mass contrast profile are superposed for the same compensation radius $R_1$ and the same halo mass.}
	\label{fig:same_mh_a}
\end{figure}

We must also ensure that modifying the simulation parameters do not affect the profiles. A numerical simulation is characterized by a mass and a spatial resolution (see Table \ref{tab:simus}). Since CoSpheres trace the matter distribution on large or intermediate scales (compared to the coarse grid size), average stacked matter profile result from the dynamics computed on the coarse grid without any refinement. As long as we consider scales larger than a few cells we should not observe any significant deviations for large scale field when changing the simulation parameters. In other words, the properties of CoSpheres are robust with respect to the resolution parameters of the simulation used to trace the matter field. We illustrate this point on \refim{fig:same_mh_a} where we plot the stacked average profile for different numerical simulations but the same halo mass $M_h = 3.0 \pm 0.075 \times 10^{13}$ \Msun ($200 \pm 5$ particles per halo) and three different compensation radii $R_1$. For each $R_1$, matter profiles are indeed merged together.

\subsubsection{The spherically compensated cosmic regions at higher redshift}
\label{sec:cosphere_initial}
CoSpheres are detected in numerical simulation at $z=0$. We then follow backward in time the evolution of the matter field of such regions using our numerical simulations. For each halo, the position of its progenitor is estimated from the center of mass of its particles at $z=0$. This estimation is correct since scales probed here are much larger than the halo size (in \refapp{shift} we show how it is possible to model a shift in the theoretical profile). For voids, \ie central minimum, we assume that its comoving position is conserved during evolution and equals to the position measured at $z=0$. For every redshift and profile, this primordial position is used to compute the spherical mass by counting the number of particles in concentric shells as discussed in section \refsec{method}. 

\begin{figure*}
	\captionsetup[subfigure]{width=0.47\linewidth}
	\begin{center}
		\subfloat[Mass contrast profile for $R_1=20$ \Mpc for various redshift from an halo with a mass $M_h = 3 \times 10^{13}$ \Msun. It is computed by stacking together single profiles computed from the progenitor of each halo registered at $z=0$. As discussed in \refsec{sec:cosphere_initial}, the position of the progenitor at any $z$ is estimated from the center of mass of the particles composing the halo at $z=0$.]{
			\includegraphics[width=0.5\textwidth]{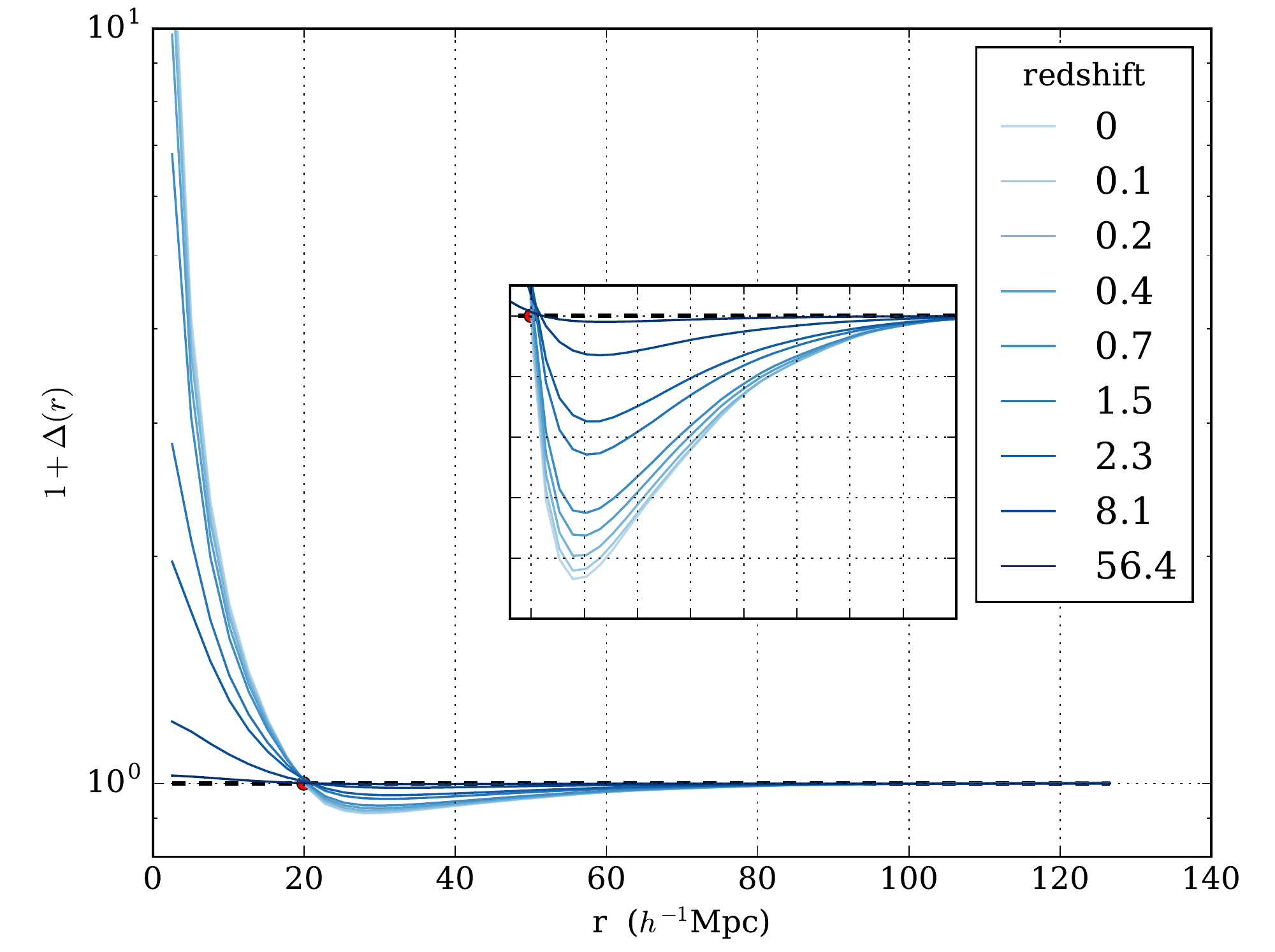}
			\label{sub:halo}}
		\subfloat[Same as in the left panel but centered on local minima and for $R_1=30$ \Mpc. For tracing backward in time the evolution of such regions, we simply assumed that the comoving positions of central minima detected at $z=0$, is conserved during the whole evolution. Such estimation provides satisfying results.]{
			\includegraphics[width=0.5\textwidth]{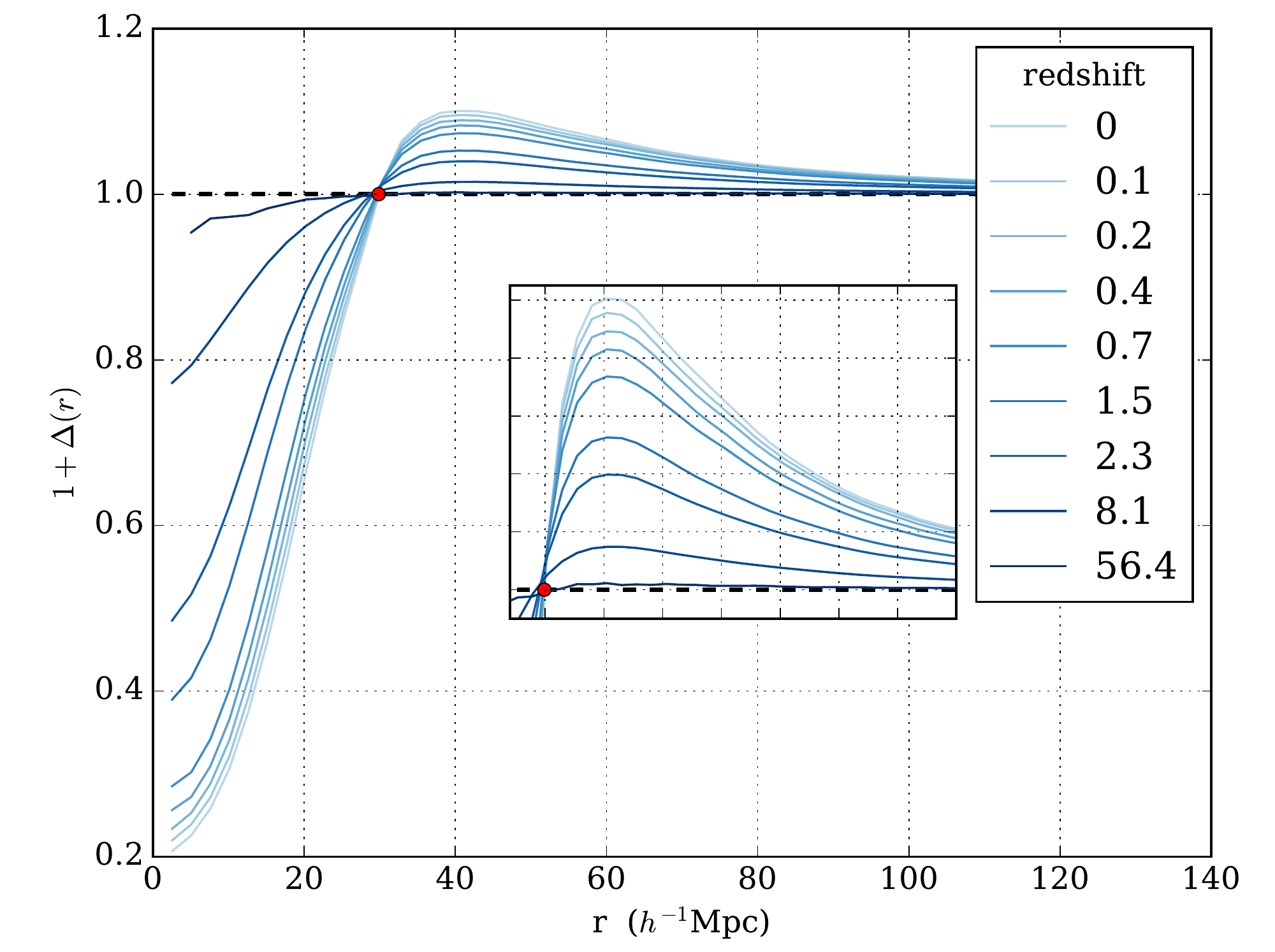}
			\label{sub:void}}
		\caption{Evolution of the mass contrast profile for both maxima and minima in the density field. The radial scale is in comoving \Mpc and each curve corresponds to a single redshift.}
		\label{fig:profile_D_evolution}
	\end{center}
\end{figure*}

\refim{fig:profile_D_evolution} shows the evolution of profiles with redshift from $z\simeq 56$ to $z=0$ and for both maxima (see \refim{sub:halo}) and minima (see \refim{sub:void}). This figure illustrates two main points. Firstly, at any redshift, the profile shows the same shape on all scales; an internal over (resp. under) massive core surrounded by its under (resp. over) massive belt. Secondly, the compensation radius (marked with a red dot at $z=0$) seems to be also conserved in comoving coordinates ($x$ axis in comoving \Mpc). This last property will be discussed later in this paper. 

\begin{figure} 	
  \centering \includegraphics[width=1.0\linewidth]{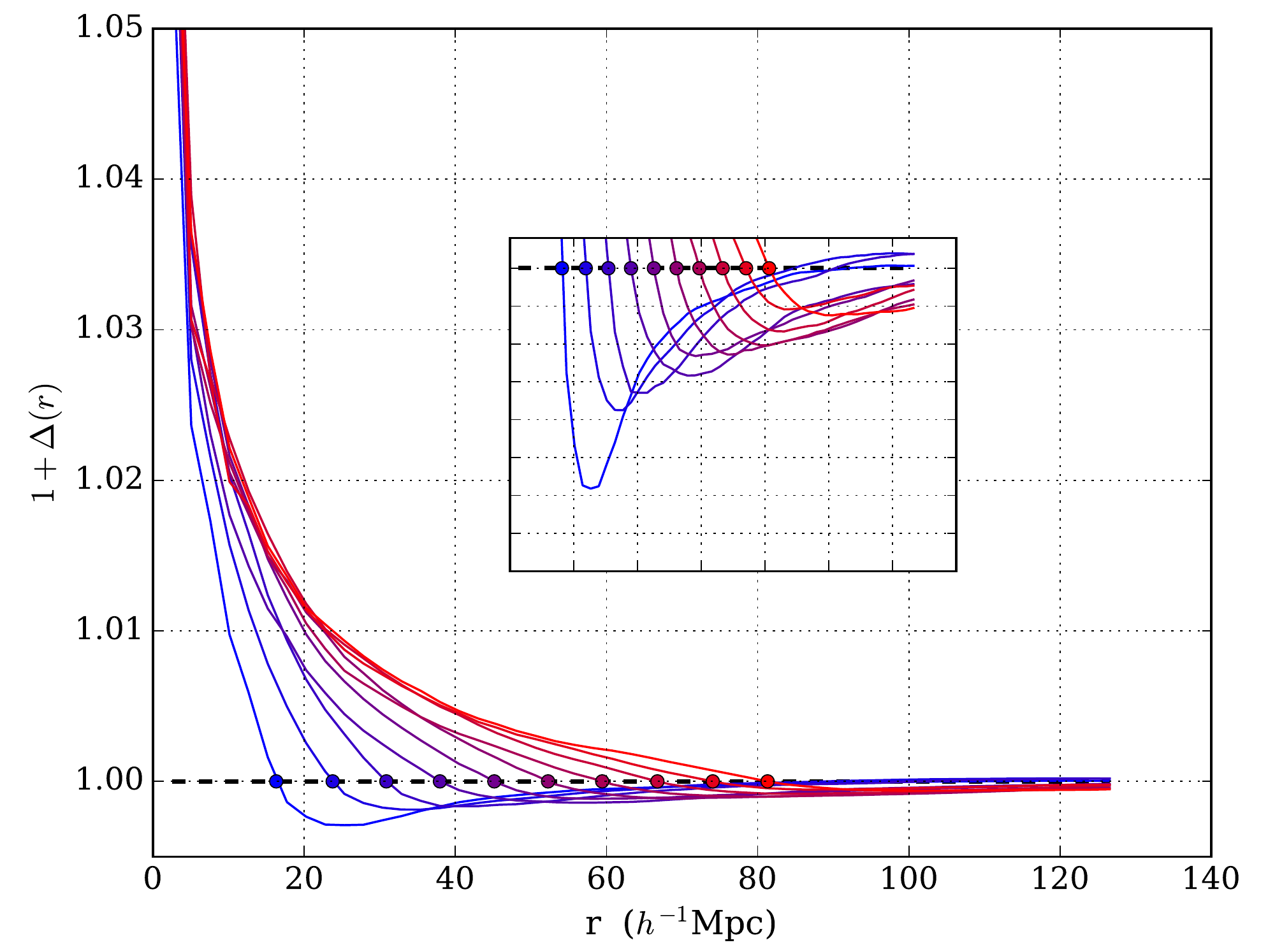}
  \caption{Mass contrast profiles at high redshift $z\simeq 57$ for different $R_1$  between $15$ and $80$ \Mpc detected originally from haloes with the same mass $M_h = 3.0\times 10^{13}$ \Msun.}
\label{fig:profile_D_initial}
\end{figure}

This indicates that CoSpheres are generated within the primordial density field at high redshift and are not generated through gravitational dynamics only. On \refim{fig:profile_D_initial}, we thus show - average - CoSphere profiles for different $R_1$ at a very high redshift $z \sim 57$ in the simulation. This figure shows that these structures are originated by large scale primordial density fluctuations with the same compensation properties. In the two following sections we will study these structures within the primordial field in the framework of GRF \refsecl{sec:initial}. The gravitational evolution of these initial profiles will be studied in \refsecl{sec:dynamic}.

\section{Origin of CoSpheres in Gaussian random field} 
\label{sec:initial}

In this section we derive the average density profile around extrema in a GRF constrained by the compensation property \refeq{R1}. Density and mass profiles of CoSpheres are characterized by two family of parameters: the peaks parameters as defined by \citep{BBKS} qualifying the central extrema and the environment parameters. 

As was studied in \citet{BBKS}, a local extrema at some position $\bx_0$ in GRF can be parametrized by $10$ independent - but correlated - parameters. A scalar $\nu$ quantifying the central height of the extrema
\begin{equation}
\label{d0}
\delta(\bold{x}_0)= \nu\sigma_0
\end{equation}
expressed in units of the fluctuation level $\sigma_0$
\begin{equation}
\sigma_0=\left[\frac{1}{2\pi^2}\int_0^\infty k^2 P(k) dk\right]^{1/2}
\end{equation}
where $P(k)$ it the linear matter power spectrum evaluated at some fixed time $t_i$ where the field can be assumed to be Gaussian (deep inside the matter-dominated era). The extremum condition imposes that the local gradient of the field $\boldsymbol{\eta}$ vanishes identically, \ie
\begin{equation}
\label{dp0}
\eta_i=\frac{\partial \delta(\bx_0)}{\partial x_i} = 0
\end{equation}
The local curvature around the extremum is described by its Hessian matrix $\boldsymbol{\zeta}$ 
\begin{equation}
\label{dpp0}
\zeta_{ij}=\frac{\partial^2 \delta(\bx_0)}{\partial x_i\partial x_j}
\end{equation}
Each eigenvalue of the Hessian matrix must be negative in the case of a central maximum and positive in the opposite case of a minimum (under-dense).

Let us now consider the environment parameters. The neighbourhood of the peak is here described by the compensation radius \refeql{R1}. However, providing this radius only is not sufficient to reconstruct the large variety of profiles and it is necessary to add the compensation density contrast defined \textit{on the sphere} of radius $R_1$. By construction it must be of opposite sign of the central density contrast. The compensation density $\delta_1$ is thus defined once averaging over angles the density on the sphere of radius $r=R_1$
\begin{equation}
\label{v1}
\delta(R_1):=\delta_1=\nu_1\sigma_0
\end{equation}
with $\nu_1/\nu<0$. Without any assumption on the symmetry, we thus need $12$ independent parameters : the scalar $\nu$, three components of the $\boldsymbol{\eta}$ vector, $6$ independent coefficients of the $\boldsymbol{\zeta}$ matrix (which is real and symmetric) together with $R_1$ and the reduced compensation density $\nu_1$. In the following, we are going to compute the expected averaged profile in the primordial Gaussian field satisfying both the peak constraints \refeq{d0}, \refeq{dp0} and \refeq{dpp0} and the environmental constraints \refeq{R1} and \refeq{v1}.

\subsection{Peaks in GRF}
\label{GRF}
Let us recall the basic elements necessary for the derivation of average quantities in the context of GRF. Our Gaussian field is assumed to be an homogeneous and isotropic random field with zeros mean. We also restrict ourselves to GRF whose statistical properties are fully determined by its power-spectrum (or spectral density) $P(k)$ \ie the Fourrier Transform of the auto-correlation function of the field, $\xi(r)=\xi(|\bold{x}_1-\bold{x}_2|)=\avg{ \delta(\bold{x}_1)\delta(\bold{x}_2)}$ :
\begin{equation}
\label{xi}
\xi(r)=\frac{1}{2\pi^2}\int_0^\infty k^2 P(k) \frac{\sin(kr)}{kr}dk
\end{equation}
The Gaussianity of the field $\delta(\bold{x})$ appears in the computation of the joint probability 
\begin{equation}
d\mathcal{P}_N=P\left[\delta(\bold{x}_1), ..., \delta(\bold{x}_N)\right]d\delta(\bold{x}_1)...d\delta(\bold{x}_N)
\end{equation}
that the field has values in the range $[\delta(\bold{x}_i), \delta(\bold{x}_i) + d\delta(\bold{x}_i)]$ for each position $\bold{x}_i$. In this particular case of homogeneous and isotropic GRF, this probability reaches
\begin{equation}
d\mathcal{P}_N=\frac{1}{\sqrt{(2\pi)^N\det \bold{M}}}\exp\left[-\frac{1}{2}\boldsymbol\delta^t.\bold{M}^{-1}.\boldsymbol\delta\right]\prod_{i=1}^Nd\delta_i
\end{equation}
where $\boldsymbol\delta$ is the $N$ dimensional vector $\delta_i=\delta(\bold{x}_i)$ and $\bold{M}$ is the $N\times N$ covariance matrix, here fully determined by the field auto-correlation
\begin{equation}
M_{ij}:=\avg{\delta_i\delta_j}=\xi(|\bold{x}_i-\bold{x}_j|)
\end{equation}
where the average operator $\avg{...}$ denotes hereafter an ensemble average. The ergodic assumption identifies the ensemble average $\avg{...}$ computed on all possible statistical realisation of the observable to its spatial averaging, \ie, its mean over a sufficiently large volume. The average of any operator $X$ can be written as a mean over its Fourier component $\tilde{X}$
\begin{equation}
\avg{X}:=\frac{1}{2\pi^2\sigma_0^2}\int_0^\infty k^2 P(k)\tilde{X}(k) dk=\frac{\int_0^\infty k^2 P(k)\tilde{X}(k) dk}{\int_0^\infty k^2 P(k) dk}
\end{equation} 
Furthermore, we are interested in deriving the properties of the field subject to $n$ linear constraints $C_1, ..., C_n$. Following \citet{Bertschinger1987}, we write each constraint $C_i$ as a convolution of the field 
\begin{equation}
\label{constraint}
C_i[\delta] := \int W_i(\bold{x}_i-\bold{x})\delta(\bold{x})d\bold{x} = c_i
\end{equation}
where $W_i$ is the corresponding window function and $c_i$ is the value of the constraint. For example, constraining the value of the field to a certain $\delta_0$ at some point $\bold{x}_0$ leads to $W_i=\delta_D(\bold{x}-\bold{x}_0)$ and $c_i=\delta_0$. Since the constraints are linear, their statistics is also Gaussian and the joint probability $d\mathcal{P}[\bold{C}]$ that the field satisfies these conditions is \citep{vandeWeygaert1996, Bertschinger1987}
\begin{equation}
d\mathcal{P}[\bold{C}] = \frac{1}{\sqrt{(2\pi)^n\det \bold{Q}}}\exp\left[-\frac{1}{2}\bold{C}^t.\bold{Q}^{-1}.\bold{C}\right]\prod_{i=1}^nd c_i
\end{equation}
where $\bold{Q}$ is the covariance matrix of the constraints $\bold{C}=\{C_1, ..., C_n \}$ defined as $\bold{Q}=\avg{\bold{C}^t.\bold{C}}$. The average density profile $\avg{\delta}$ subject to $\bold{C}$ is computed as the most probable profile given $\boldsymbol{C}$ and reaches
\begin{equation}
\label{average}
\avg{\delta}(\bold{x}):=\avg{\delta|\bold{C}}(\bold{x})= \xi_i(\bold{x})Q_{ij}^{-1}c_j
\end{equation}
where $\xi_i(\bold{x})$ is the correlation function between the field and the $i^\text{th}$ constraint 
\begin{equation}
\label{correlations}
\xi_i(\bold{x})=\avg{\delta(\bold{x})C_i}
\end{equation}
and $Q_{ij}$ is the $(ij)$ element of the correlation matrix $\bold{Q}$. \cite{BBKS} derived the average spherical\footnote{their calculation goes beyond the spherical approximation but we restrict here to the simplest spherical case obtained by averaging over angles} density profile of peaks in GRF in term of the reduced height $\nu$ \refeql{d0}, its auto-correlation function $\xi(r)$ \refeql{xi} and the curvature parameter $x$ defined by
\begin{equation}
\label{x_def}
x = -\frac{3}{\sigma_0\sqrt{\avg{k^4}}}\frac{\partial^2\delta (\bold{x}_0)}{\partial r^2}
\end{equation}
It yields
\begin{equation}
\label{bbks}
\avg{\delta}_{peaks}(r)=\frac{\nu-\gamma x}{1-\gamma^2}\frac{\xi(r)}{\sigma_0}+\frac{\nu-x/\gamma}{1-\gamma^2}\frac{R_\star^2}{3}\frac{\Delta \xi(r)}{\sigma_0}
\end{equation}
with $R_\star=\sqrt{3\avg{k^2}/\avg{k^4}}$ and $\gamma = \avg{k^2}/\sqrt{\avg{k^4}}$. The various moments of $k$ are given by 
\begin{equation}
\avg{k^n}:=\frac{1}{2\pi^2\sigma_0^2}\int_0^\infty k^{2+n} P(k) dk
\end{equation} 
In the next section, we extend this result by implementing the compensation conditions \refeq{R1} and \refeq{v1}.

\subsection{CoSpheres in Gaussian Random Fields}
\label{new parametrisation}

In the following, we use several functions involving $r$ and $R_1$. In order to simplify the notations, we note the Fourier components as 
\begin{align}
W_r &:= 3\frac{\sin(kr)-kr\cos(kr)}{(kr)^3}\\
J_r &:= \frac{\sin(kr)}{kr}
\end{align}
Which are respectively the Fourier transform of the top-hat and the delta functions. Note that they are linked to the spherical Bessel functions as $W_r=3/(kr)j_1(kr)$ and $J_r=j_0(kr)$. We also denote with the "1" subscript these quantities evaluated at the particular radius $r=R_1$, \ie
\begin{align}
W_1 &:= \left.W_r\right|_{r=R_1}\\
J_1 &:= \left.J_r\right|_{r=R_1}
\end{align}

\subsubsection{The average density profile of CoSphere}

We now derive the average matter profile being both
\begin{enumerate}
	\item centred on an extremum, \ie satisfying the conditions \refeqs{d0}{dp0}{dpp0}
	\item and compensated on a finite scale $R_1$. This is implemented by the compensation constraints \refeq{R1} and \refeq{v1}.
\end{enumerate}
The spherical peak constraints are \refeql{constraint}
\begin{align}
C_{\nu}[\delta]&:=\int \delta_D\left[\bold{x}-\bold{x}_0\right]\delta(\bold{x})d\bold{x}=c_{\nu}\equiv \sigma_0\nu\\
C_{\eta_i}[\delta]&:=\int \partial_i\delta_D\left[\bold{x}-\bold{x}_0\right]\delta(\bold{x})d\bold{x}=c_{\eta_i}\equiv 0\\
C_{x}[\delta]&:=\int \frac{\partial^2}{\partial \bold{x}^2}\delta_D\left[\bold{x}-\bold{x}_0\right]\delta(\bold{x})d\bold{x}=c_x\equiv -\sigma_0\frac{x\sqrt{\avg{k^4}}}{3}
\end{align}
where $\delta_D$ is the usual Dirac function. On the other hand, the \textit{environmental} constraints (see \refeq{R1} and \refeq{v1}) can be written
\begin{align}
\label{R1_constraint}
C_{R_1}[\delta]&:=\int \Theta\left[R_1-|\bold{x}-\bold{x}_0|\right]\delta(\bold{x})d\bold{x}=c_{R_1} \equiv 0\\
C_{\nu_1}[\delta]&:=\int \delta_D\left[R_1-|\bold{x}-\bold{x}_0|\right]\delta(\bold{x})d\bold{x}=c_{\nu_1}\equiv \sigma_0\nu_1
\end{align}
where $\Theta$ is the Heaviside step function. $\nu_1$ must also satisfy \refeq{v1} and thus satisfy $\nu_1/\nu<0$. Note that in \refeq{R1_constraint}, the constraint value is $c_{R_1}=0$ and only three eigenvalues for the constraints are non zero : $\nu$, $x$ and $\nu_1$. The original peak constraints involve the correlations
\begin{align}
\label{xi_v}
\xi_{\nu}(r)&:=\avg{\delta(\bold{x})C_{\nu}}=\sigma_0^2\avg{J_r}\\
\xi_{x}(r)&:=\avg{\delta(\bold{x})C_{x}}=-\sigma_0^2\avg{k^2J_r}
\end{align}
The compensation constraints introduce new correlations in the computation of the average profile
\begin{align}
\xi_{R_1}(r)&:=\avg{\delta(\bold{x})C_{R_1}}=\sigma_0^2\avg{W_1J_r}\\
\label{xi_v1}
\xi_{\nu_1}(r)&:=\avg{\delta(\bold{x})C_{\nu_1}}=\sigma_0^2\avg{J_1J_r}
\end{align}
Using \refeq{average}, the average profiles are linear in $\nu$, $x$ and $\nu_1$ while $R_1$ implicitly appears in the various radial functions, such that we can write
\begin{equation}
\label{profile_0}
\boxed{\avg{\delta}(r)=\sigma_0\Big(\nu\delta_\nu(r) + x\delta_x(r) + \nu_1\delta_{\nu_1}(r)\Big)}
\end{equation}
where $\delta_\alpha$ with  $\alpha=\{\nu,x,\nu_1\}$ are functions of $r$ and $R_1$ only. By construction, each $\delta_\alpha(r)$ must satisfy the compensation property, \ie its integral must vanish at $r=R_1$
\begin{equation}
\label{comp}
\int_0^{R_1}u^2\delta_\alpha(u)du=0
\end{equation}
From \refeq{average} we know that each $\delta_\alpha(r)$ is a linear combination of the four functions $\xi_{\alpha}(r)$ (see \refeq{xi_v} to \refeq{xi_v1}). To simplify the notations, we define three intermediate functions build from the $\xi_{\alpha}$ functions and satisfying the condition \refeq{comp}
\begin{align}
\label{xi0}
f_0(r)&=\frac{\avg{k^2W_1}\avg{J_r}-\avg{W_1}\avg{k^2J_r}}{\avg{k^2W_1}-\avg{k^2}\avg{W_1}}\\
f_1(r)&=\frac{\avg{W_1J_r}\avg{k^2W_1}-\avg{W_1^2}\avg{k^2J_r}}{\avg{W_1}\avg{k^2W_1}-\avg{k^2}\avg{W_1^2}}\\
f_2(r)&=\frac{\avg{J_1J_r}\avg{k^2W_1}-\avg{J_1W_1}\avg{k^2J_r}}{\avg{J_1}\avg{k^2W_1}-\avg{k^2}\avg{J_1W_1}}
\end{align}
these functions have also been normalized such that $f_i(0)=1$. Each $\delta_\alpha(r)$ is then a linear combination of these three functions. Note that $f_i(R_1)\neq 0$. We also introduce three parameters $\lambda_i$ defined locally around the extremum 
\begin{equation}
\lambda_i:=-3\frac{\partial^2f_i}{\partial r^2}\quad\text{ for}\quad r\to 0
\end{equation}
They explicitly reach 
\begin{align}
\lambda_0&=\frac{\avg{k^2W_1}\avg{k^2}-\avg{W_1}\avg{k^4}}{\avg{k^2W_1}-\avg{k^2}\avg{W_1}}\\
\lambda_1&=\frac{\avg{k^2W_1}^2-\avg{W_1^2}\avg{k^4}}{\avg{W_1}\avg{k^2W_1}-\avg{k^2}\avg{W_1^2}}\\
\lambda_2&=\frac{\avg{k^2J_1}\avg{k^2W_1}-\avg{J_1W_1}\avg{k^4}}{\avg{J_1}\avg{k^2W_1}-\avg{k^2}\avg{J_1W_1}}
\end{align}
With these notations and a bit of algebra we obtain the $\delta_\alpha$ functions
\begin{align}
\label{delta_nu}
\nonumber \delta_\nu(r)=f_0(r)\frac{\lambda_1f_2^1 - \lambda_2f_1^1}{\omega} &+ f_1(r)\frac{\lambda_2f_0^1 - \lambda_0f_2^1}{\omega} \\& +f_2(r)\frac{\lambda_0f_1^1-\lambda_1f_0^1}{\omega}
\end{align}
together with
\begin{equation}
\label{delta_x}
\frac{\delta_x(r)}{\sqrt{\avg{k^4}}}=f_0(r)\frac{f_1^1-f_2^1}{\omega}+f_1(r)\frac{f_2^1-f_0^1}{\omega} + f_2(r)\frac{f_0^1-f_1^1}{\omega}
\end{equation}
and 
\begin{equation}
\label{delta_nu1}
\delta_{\nu_1}(r)=f_0(r)\frac{\lambda_2-\lambda_1}{\omega} + f_1(r)\frac{\lambda_0-\lambda_2}{\omega}+f_2(r)\frac{\lambda_1-\lambda_0}{\omega}
\end{equation}
where
\begin{equation}
\omega=\lambda_0\left(f_1^1-f_2^1\right)+\lambda_1\left(f_2^1-f_0^1\right)+ \lambda_2\left(f_0^1-f_1^1\right)
\end{equation}
and we used the notation $f_i^1:=f_i(R_1)$. Note that $\lambda_i$ and $f_i^j$ are not constant but non linear functions of $R_1$. 

The $\delta_\alpha$ functions satisfy the following properties around $r=0$
\begin{equation}
\begin{cases}
\delta_\nu(r)&\simeq 1 +\order{r^4}\\
\delta_x(r)&\simeq -\sqrt{\avg{k^4}}\frac{r^2}{6}+\order{r^4}\\
\delta_{\nu_1}(r)&\simeq \order{r^4}
\end{cases}
\end{equation}
while in $r=R_1$ we have
\begin{equation}
\begin{cases}
\delta_{\nu}(R_1)&=0\\
\delta_x(R_1)&=0\\
\delta_{\nu_1}(R_1)&=1
\end{cases}
\end{equation}
insuring that $\avg{\delta}(R_1)=\sigma_0\nu_1$.

\begin{figure*}
	\includegraphics[width=1.0\linewidth]{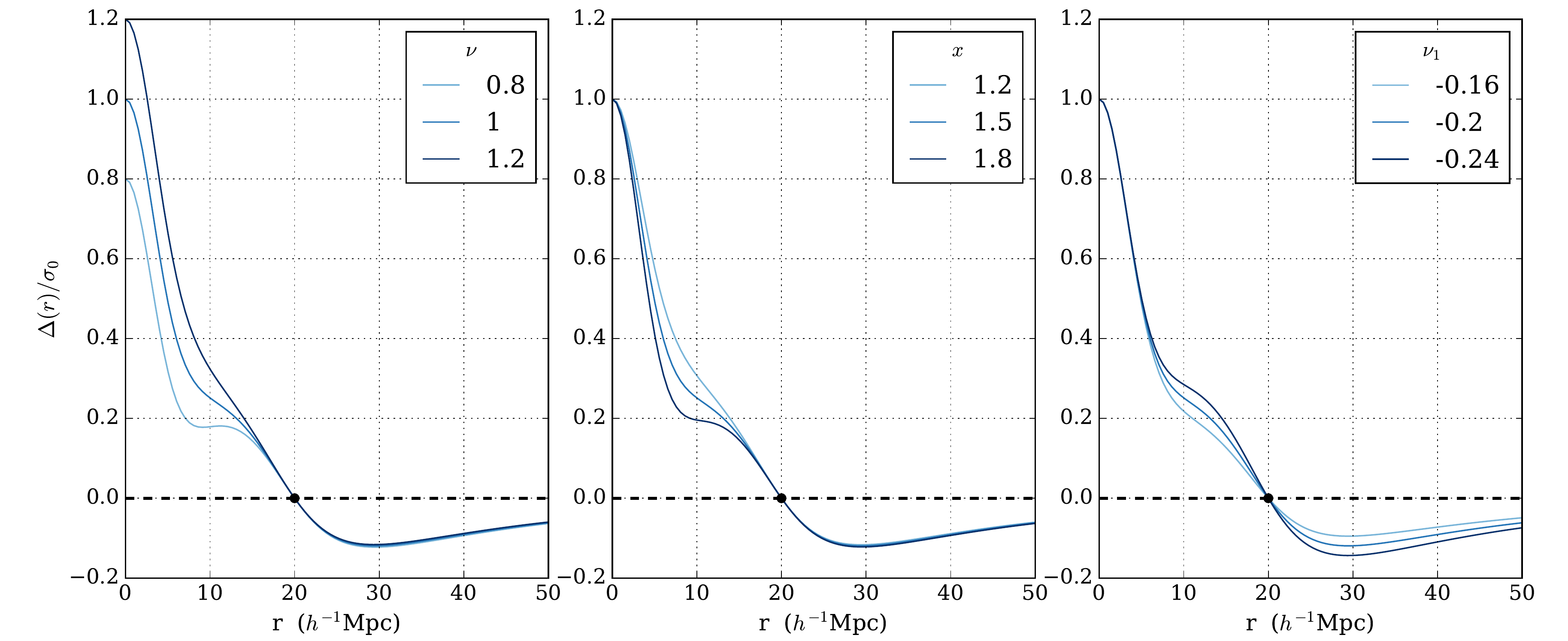}
	\caption{Rescaled mass contrast profile ($\Delta(r)/\sigma_0$) derived in \refeq{Delta_i} for $R_1=20$ \Mpc (black dot). On each panel we vary one of the shape parameters $\nu$, $x$ or $\nu_1$. This figure illustrates that the peak parameters $\nu$ and $x$ affect the profile on small scales (around the peak) while $\nu_1$ acts on the large scale shape of the matter profile. Note that $\nu_1$ changes the slope of the mass contrast profile around $R_1$ since by definition $\Delta'(R_1)=3/R_1\sigma_0 \nu_1$. This analysis is absolutely symmetric for the case of a central under-density. Note that the elbow around $r=10$ \Mpc is an artificial feature induced by forcing one shape parameter to vary while fixing the other ones. In practice, these shape parameters are correlated together and this scale does not appears in realistic profiles (see other figures for example). We discuss these correlations in more detail in \citet{paper2}.}
	\label{fig:profile_D_parameters}
\end{figure*}

\subsubsection{The averaged mass contrast profile of CoSphere}
\label{averagedcompensated}

The average mass contrast profile $\avg{\Delta}(r)$ is obtained by integrating $\avg{\delta}(r)$ with \refeq{Delta0}. By linearity of the mapping $\delta\Leftrightarrow \Delta$, $\avg{\Delta}$ takes the same shape than \refeq{profile_0} where each $\delta_\alpha(r)$ transforms to $\Delta_\alpha(r)$, \ie we have
\begin{equation}
\label{Delta_i}
\boxed{\avg{\Delta}(r)=\sigma_0\Big(\nu\Delta_\nu(r) + x\Delta_x(r) + \nu_1 \Delta_{\nu_1}(r)\Big)}
\end{equation}
Since each function $\delta_\alpha$ is a linear combination of the $f_i$, the $\Delta_\alpha$ functions will be linear combinations of the $F_i$ functions defined as
\begin{equation}
F_i(r):=\frac{3}{r^3}\int_0^ru^2f_i(u)du
\end{equation}
Moreover, the $f_i$ functions only involve linear combinations of $J_r=\sin(kr)/(kr)$. The $F_i$ are thus obtained from $f_i$ by the simple replacement $J_r \to W_r$, leading to
\begin{align}
F_0(r)&=\frac{\avg{k^2W_1}\avg{W_r}-\avg{W_1}\avg{k^2W_r}}{\avg{k^2W_1}-\avg{k^2}\avg{W_1}}\\
F_1(r)&=\frac{\avg{W_1W_r}\avg{k^2W_1}-\avg{W_1^2}\avg{k^2W_r}}{\avg{W_1}\avg{k^2W_1}-\avg{k^2}\avg{W_1^2}}\\
F_2(r)&=\frac{\avg{J_1W_r}\avg{k^2W_1}-\avg{J_1W_1}\avg{k^2W_r}}{\avg{J_1}\avg{k^2W_1}-\avg{k^2}\avg{J_1W_1}}
\end{align}
We can check that for $i=\{0 ,1, 2\}$ we have $F_i(R_1)=0$ insuring that $\avg{\Delta}(R_1)=0$ whatever the shape parameters. The mapping between the $\Delta_\alpha$ functions and the $F_i$ is given by
\begin{align}
\label{Delta_nu}
\nonumber \Delta_\nu(r)=F_0(r)\frac{\lambda_1f_2^1 - \lambda_2f_1^1}{\omega} &+ F_1(r)\frac{\lambda_2f_0^1 - \lambda_0f_2^1}{\omega} \\& +F_2(r)\frac{\lambda_0f_1^1-\lambda_1f_0^1}{\omega}
\end{align}
while for $x$ we have
\begin{equation}
\label{Delta_x}
\frac{\Delta_x(r)}{\sqrt{\avg{k^4}}}=F_0(r)\frac{f_1^1-f_2^1}{\omega}+F_1(r)\frac{f_2^1-f_0^1}{\omega} + F_2(r)\frac{f_0^1-f_1^1}{\omega}
\end{equation}
and 
\begin{equation}
\label{Delta_nu1}
\Delta_{\nu_1}(r)=F_0(r)\frac{\lambda_2-\lambda_1}{\omega} + F_1(r)\frac{\lambda_0-\lambda_2}{\omega}+F_2(r)\frac{\lambda_1-\lambda_0}{\omega}
\end{equation}
The resulting mass contrast profile satisfies, for $r\to 0$
\begin{equation}
\begin{cases}
\Delta_\nu(r)&\simeq 1 +\order{r^4}\\
\Delta_x(r)&\simeq -\sqrt{\avg{k^4}}\frac{r^2}{10}+\order{r^4}\\
\Delta_{\nu_1}(r)&\simeq \order{r^4}
\end{cases}
\end{equation}
while in $r=R_1$ we have, by construction
\begin{equation}
\begin{cases}
\Delta_{\nu}(R_1)&=0\\
\Delta_x(R_1)&=0\\
\Delta_{\nu_1}(R_1)&=0
\end{cases}
\end{equation}
together with
\begin{equation}
\Delta_{\nu_1}'(R_1)=\frac{3}{R_1}\nu_1\sigma_0
\end{equation}
where a prime denotes the derivative with respect to $r$. On \refim{fig:profile_D_parameters} we plot the averaged mass contrast \refeq{Delta_i} for $R_1=20$ \Mpc. In each panel we change one of the shape parameters $\nu$, $x$ and $\nu_1$ to illustrate their effect. Since both $\nu$ and $x$ are associated with the central peak, changing their value only affects the profile on small scales, typically $r \lesssim R_1/2$. The compensation density $\nu_1$ defines the structure of the profile on larger scales from $r\sim R_1$. This behaviour clearly illustrates that $x$ and $\nu$ are defined on the peak while $R_1$ and $\nu_1$ qualify the large scale surrounding environment of the peak.
 
\subsubsection{Comparison with the BBKS peak profile}

The peak profiles derived by BBKS \refeql{bbks} describe the large scale environment surrounding extrema in Gaussian Field where we only provide the properties of the density field on the peak. Our calculation is thus an extension of this result including the physical properties of the large scale environment around the peak.

Our formalism allows to probe different cosmic environment for the same central extremum. This environment is defined through $R_1$ and the compensation density $\delta_1=\nu_1 \sigma_0$. For the same central peak, we can describe a large variety of cosmic configurations while this region is completely fixed in the standard BBKS approach.

\begin{figure}
	\includegraphics[width=1.0\linewidth]{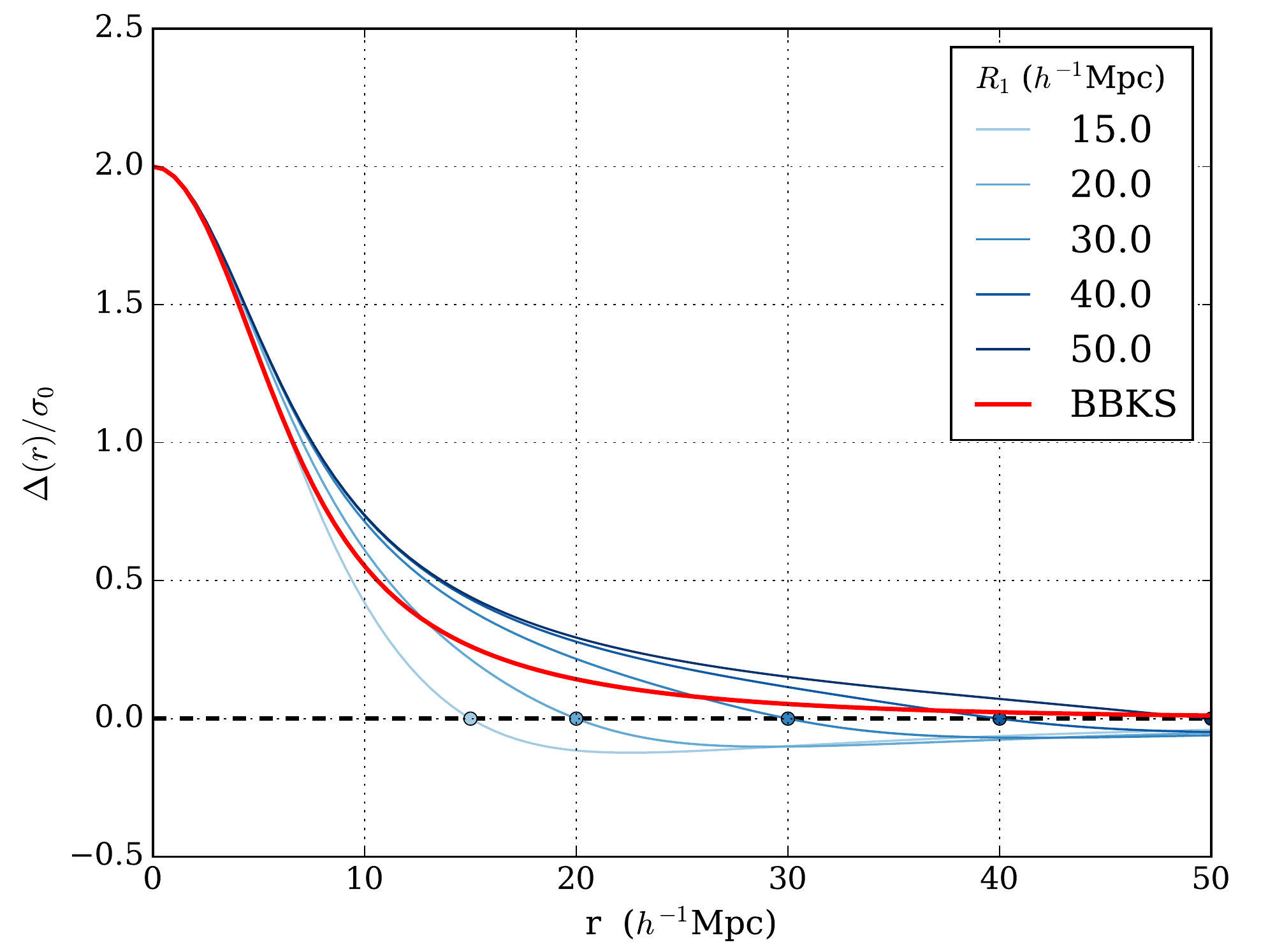}
	\caption{Averaged mass contrast profile in the primordial field. Each blue curve corresponds to a CoSphere profile with fixed peak parameters ($x$ and $\nu$) but various $R_1$ ($\nu_1$ is also fixed to an arbitrary value of $-0.5$). The red curve is the corresponding BBKS profile \refeq{bbks} with the same peak parameters $\nu$ and $x$. Local matter profile around the extremum are similar while on larger scales our model allow to probe different cosmic environments through $R_1$ and $\nu_1$. Even if it does not appears on the plot, this BBKS profile is also compensated on a larger scale, here $R_1\simeq 60$ \Mpc.}
	\label{fig:profile_D_bbks_comparison}
\end{figure}

In \refim{fig:profile_D_bbks_comparison} we show how it is possible to describe various environments by varying $R_1$ while keeping constant $\nu$ and $x$, \ie the central peak. We also plot the BBKS profile, fully determined by $x$ and $\nu$. Small values of $R_1$ correspond to isolated peaks in large under dense regions while increasing the compensation scale allows to probe denser regions. The exact same symmetric case occurs for cosmic voids with the reimplement $\nu\to -\nu$, $x\to - x$ and $\nu_1 \to -\nu_1$.

The standard BBKS profile can be written as in \refeq{profile_0} with the corresponding $\delta_\alpha^p(r)$ (where $p$ stands for "peak")
\begin{align}
\delta_\nu^p(r)=&\frac{\avg{k^4}\avg{J_r}-\avg{k^2}\avg{k^2J_r}}{\avg{k^4}- \avg{k^2}^2}\\
\delta_x^{p}(r)=&-\sqrt{\avg{k^4}}\left[\frac{\avg{k^2}\avg{J_r}-\avg{k^2J_r}}{\avg{k^4}- \avg{k^2}^2}\right]\\
\delta_{\nu_1}^{p}(r)=&0
\end{align}
In the peak profile of BBKS, it is possible to map the peak parameters $\nu$ and $x$ to their effective $R_1$ and $\nu_1$. These effective $R_1$ and $\nu_1$ satisfy
\begin{equation}
\begin{cases}
\lambda_0(R_1)&=\sqrt{\avg{k^4}}x/\nu\\
\nu_1&=\nu f_0^1(R_1)
\end{cases}
\end{equation}
where we recall
\begin{align}
\lambda_0&=\frac{\avg{k^2W_1}\avg{k^2}-\avg{W_1}\avg{k^4}}{\avg{k^2W_1}-\avg{k^2}\avg{W_1}}\\
f_0^1&=\frac{\avg{k^2W_1}\avg{J_1}-\avg{W_1}\avg{k^2J_1}}{\avg{k^2W_1}-\avg{k^2}\avg{W_1}}
\end{align}
Note that this effective compensation radius $R_1$ depends only on the fraction $x/\nu$. For each value of $x/\nu$ it is possible to find a finite compensation radius $R_1$.

\subsection{Numerical Reconstruction of CoSpheres in GRF}
\label{sec:initial_reconstruction}

\begin{figure}
    \includegraphics[width=0.5\textwidth]{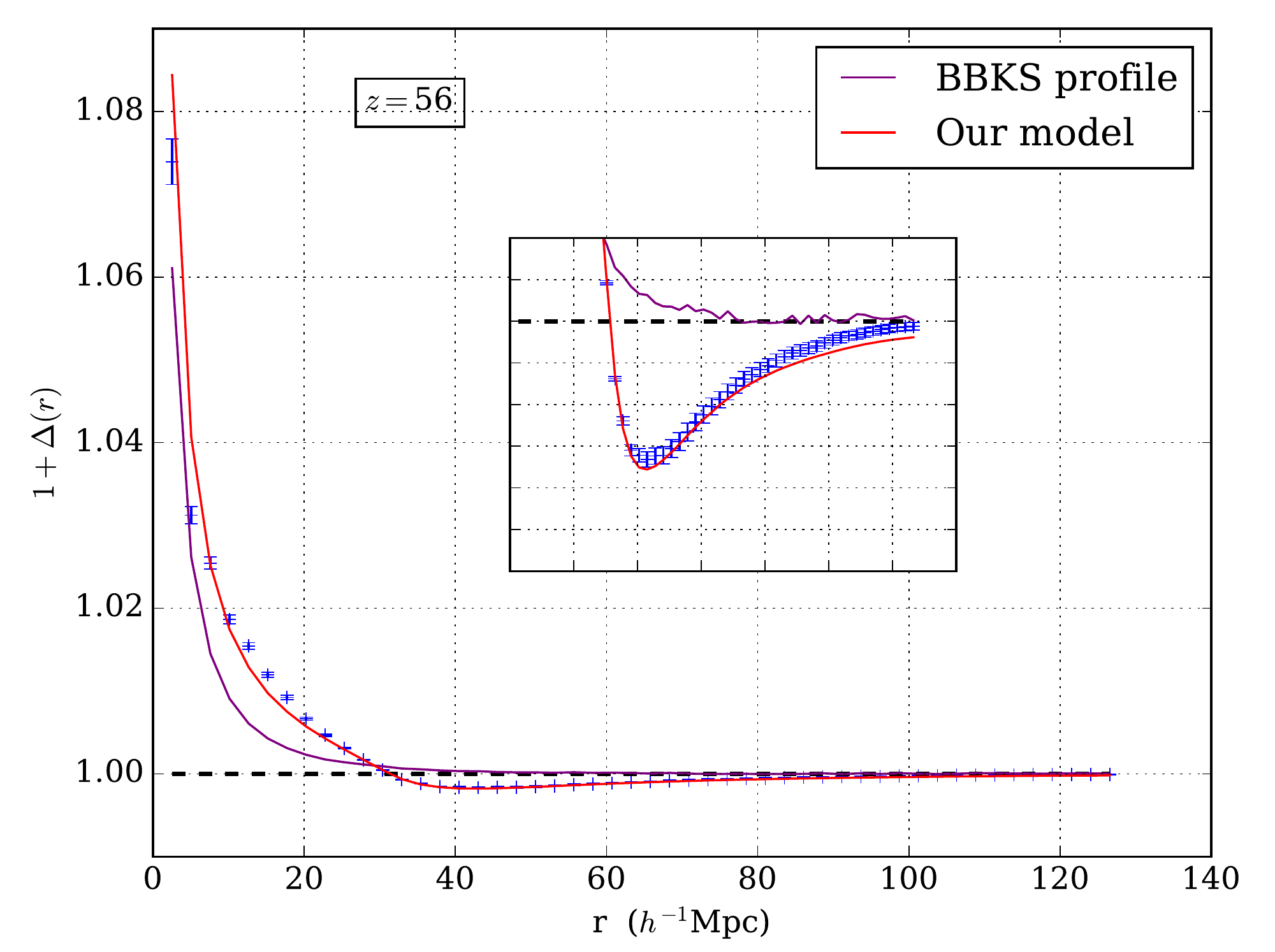}
    \caption{Averaged mass contrast profile at high redshift $z=56$ in the reference simulation (blue curve). The purple curve is the standard BBKS profile \citep{BBKS} with $\nu$ and $x$ obtained by a standard $\Chi^2$ minimization. The red curve is the CoSphere profile \refeq{Delta_i} whose shape parameters ($\nu$, $x$ and $\nu_1$) are obtained by the same $\Chi^2$ optimization. We note the excellent agreement between our profile and the measured one while the standard peak formalism fails to reproduce the shape on the matter field.}
    \label{fig:initial_profiles}
\end{figure}

At very high redshift the density field follows a Gaussian statistics. This property is inherited from the inflation phase of the young Universe. In the previous sections we derived the average profiles of spherically compensated inhomogeneities in the framework of GRF. As presented in \refsec{sec:cosphere_initial}, the simulations can be used to follow backward in time the evolution of CoSpheres detected at $z=0$. Using this numerical procedure we can compare our expected Gaussian profiles \refeql{profile_0} with the numerical matter field of compensated peaks at higher redshift.

Each theoretical profile is parametrized by four scalars. The compensation radius $R_1$ can be read directly from the profile (build at fixed $R_1$). The shape parameters $\nu$, $x$ and $\nu_1$ are computed using a standard $\Chi^2$ method defined from the measured profile $\Delta_j$ and its error $\sigma_j$ for $r=r_j$.

We stress that this reconstruction is done on the mass contrast profile and not directly from the density profile. Indeed, computing the number of particles in concentric spheres leads naturally to the mass contrast while the density must be computed by taking its radial derivative \refeql{structural_relation}. Moreover, since mass contrast is integrated over $r$, its statistical noise is smaller than for $\delta(r)$.

At high redshift, we allow a possible shift $R=\abs{\boldsymbol{x}_c- \boldsymbol{x}_0}$ between the true position of the central extrema $\boldsymbol{x}_0$ and its estimation $\boldsymbol{x}_c$ by adding this degree of freedom in the previous $\Chi^2$ analysis. The effect of such shift of the central position on theoretical profiles is discussed in \refapp{shift}. We show that the analytical profile around a shifted position is modified by an effective smoothing of the linear power spectrum $P(k)\to P(k)\sinc(kR)$. 

In \refim{fig:initial_profiles} we show a measured mass contrast profile at high redshift (blue curve). We show in red the corresponding expected profile \refsecl{averagedcompensated} whose shape parameters are estimated from the previous $\Chi^2$. We also show in purple the expected shape from the BBKS profile with best fitted values for $x$ and $\nu$

We observe a high accuracy agreement between our prediction \refeq{Delta_i} and the measured averaged profile on all scales (this fit can be reproduce for any $R_1$). The standard BBKS profile suffers from a lack of degrees of freedom and is not able to reproduce the shape on large scales. The standard peak profile cannot be used to describe the large scale compensated cosmic regions, hence the CoSphere model.

\FloatBarrier
\section{Dynamical evolution of CoSpheres}
\label{sec:dynamic}
In this section we study the gravitational collapse of CoSpheres resulting from the primordial fluctuations of the matter field as studied in \refsec{sec:initial}.

\subsection{The Lagrangian Spherical Collapse}
\label{sec:SC}
Due to the spherical symmetry of our problem, the evolution of the matter profile reduces to the dynamics of concentric shells with fixed mass. This leads to the Lagrangian spherical collapse model, first introduced in \citet{Gunn1972} and largely discussed in \citep{Padmanabhan1993,Peacock1998}. While it was first developed in the context of Einstein-de-Sitter cosmology, it has been extended to $\Lambda$CDM \citep{Lahav1991} and more general models of Dark Energy \citep{Wang1998}. In this section we derive a simple formalism for the spherical collapse suited to our problem.

As we focus in this paper on the standard $\Lambda$CDM model describing a flat Universe ($K=0$) with collision-less Cold Dark Matter and a cosmological constant $\Lambda$, the homogeneous background is described by the Friedman and the Raychaudhuri equations 
\begin{align}
\label{back0}
&\left(\frac{\dot{a}}{a}\right)^2 = H_0^2\left[\frac{\Omega_{m}^0 }{a^3} + 1-\Omega_{m}^0\right]\\
\label{back00}
&\frac{\ddot{a}}{a} = -\frac{H_0^2}{2}\left[\frac{\Omega_{m}^0}{a^3} +2\Omega_{m}^0-2\right]
\end{align}
where a dot ($e.g.$ $\dot{X}$) denotes the derivative with respect to the proper time $t$, $H_0$ is the Hubble constant today and  $\Omega_{m}^0 = 8\pi G \bar{\rho}_{m,0}/(3H_0^2)$. Moreover, in the Quasi Static Limit (QSL) where the time variation of the gravitational potential are smooth, \ie $\dot{\Phi} \ll \Phi \dot{a}/a$ and for scales deep inside the Hubble radius $r \ll 1/(aH)$, the first order perturbed equations reduce to the well known Poisson equation linking the local density contrast $\delta$ to the Newtonian potential $\Phi$

\begin{equation}
\label{poisson}
\nabla^2\Phi = 4\pi G\bar{\rho}_m\delta
\end{equation}
Using the spherical symmetry, the Poisson equation can be integrated once to give the Newtonian acceleration
\begin{equation}
\label{poisson2}
\frac{\partial\Phi}{\partial r}=r\frac{4\pi G\bar{\rho}_m}{3}\Delta(r)
\end{equation}
with $\Delta(r)$ is the mass contrast. It thus drives the local gravitational acceleration. Note that for $r=R_1$ we have $\grad{\Phi}=0$.
In the QSL, the equation of motion of each shell with a physical radius $r=a\Chi$ is \citep{Peebles1980}
\begin{equation}
\label{eom2}
\ddot{r} = \frac{\ddot{a}}{a}r -\grad{\Phi}
\end{equation}
For each shell we define the dimensionless comoving displacement
\begin{equation}
\label{normalisation}
\varsc(\Chi_i, t)=\frac{\Chi(t)}{\Chi_i}
\end{equation}
where $\Chi(t)$ is the comoving radius of the shell at some time $t$ and $\Chi$ its initial radius $\Chi_i=\Chi(t_i)$. The equation of motion for each concentric shell can be simplified assuming there is no \textit{shell-crossing} (we discuss this hypothesis below) insuring the mass conservation
\begin{equation}
\label{fnoSC}
\frac{1+\Delta}{1+\Delta_i}=\varsc^{-3}
\end{equation}
where $\Delta_i$ is the initial mass contrast profile of the Lagrangian shell $\Delta_i=\Delta\big(\Chi(t_i), t_i\big)$ while $\Delta$ is the evolved mass contrast $\Delta=\Delta(\Chi, t)$. We also introduce the logarithmic scale factor $\vart$ defined through 
\begin{equation}
\label{tildeaplha}
\frac{d \vart}{d\log(a)}:=\sqrt{\frac{\Omega_m}{2}}
\end{equation}
For $\Lambda$CDM, assuming $\vart(a_i)=0$ we have
\begin{equation}
\label{theta}
\vart(a)=\frac{\sqrt{2}}{3}\left[\arctanh\left(\Omega_{m,i}^{-1/2}\right)-\arctanh\left(\Omega_{m}^{-1/2}\right)\right]
\end{equation}
With this new parametrisation, the equation of motion for each concentric shell \refeql{eom2} reaches
\begin{equation}
\label{dynamic}
\boxed{\frac{\partial^2 \varsc}{\partial\vart^2}+\frac{1}{\sqrt{2\Omega_m}}\frac{\partial\varsc}{\partial\vart} = \varsc-\frac{1+\Delta_i}{\varsc^2}}
\end{equation}
\refeq{dynamic} describes the non-linear gravitational evolution of each shell until shell-crossing. Even if it does not appears now, the formulation \refeq{dynamic} can be simply extended to any cosmologies including extensions of Gravity as we will show it in forthcoming papers \citep{paper3, paper4}. The initial conditions for this differential problem are given by 
\begin{equation}
\label{ini0}
\varsc(t_i)=1
\end{equation}
together with the first derivative $\partial_\vart\varsc(t_i)$. It can be estimated from the high redshift solution where the field follows the Zel'dovich dynamic (see \refapp{sec:Zeldo})
\begin{equation}
\label{ini2}
\frac{\partial \varsc}{\partial \vart}(t_i)=-\left.\sqrt{\frac{2}{\Omega_{m,i}}}\frac{\Delta_i}{3}\frac{d\log(D)}{d\log(a)}\right\vert_{t_i}:=-\sqrt{\frac{2}{\Omega_{m,i}}}\frac{\Delta_i}{3}f(t_i)
\end{equation}
where $f(t_i)$ is the linear growth rate evaluated at the initial time $t_i$. The non linearly evolved profile $\Delta$ is obtained by solving numerically \refeq{dynamic} and using \refeq{fnoSC} for any initial profile $\Delta_i$.

\begin{figure}
	\includegraphics[scale=0.45]{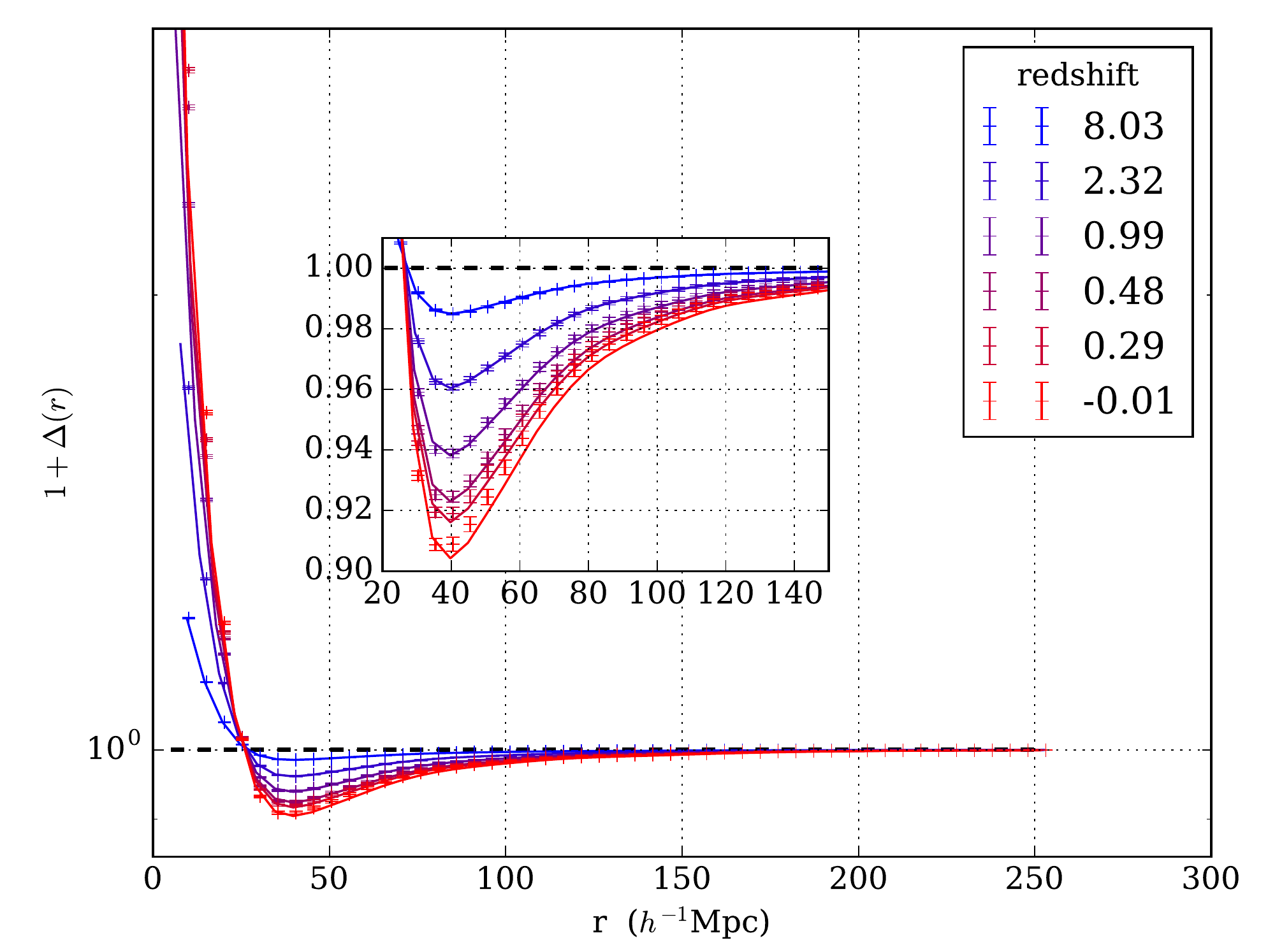}
	\caption{Comparison between average numerical profiles and the spherical evolution of the corresponding average primordial numerical matter profile for $z\simeq 8$ to $z=0$ (see text). Points and their associated error bars are the numerical measures for the corresponding redshift while full lines are the spherical evolution of the primordial profile at $z\simeq 8$. The computation has been performed using the $2048^3$ particles simulation with a box size of $5184$ \Mpc in $\Lambda$CDM cosmology and for $R_1=25$ \Mpc. All single profiles have been detected from central haloes of mass $M_h\simeq 2.50\pm 0.13 \times 10^{14}$ \Msun at $z=0$.} 
	\label{fig:spherical_collapse}
\end{figure}

\subsection{Testing the spherical approximation for the evolution of CoSpheres}
\label{sec:spherical_dynamics}

The validity of the spherical evolution can be tested using the numerical simulations. 

At $z=0$ we select haloes with the same $R_1$. For each halo detected we apply the "backward" procedure discussed in \refsec{sec:cosphere_initial} to build the profile of its progenitor. Stacking these primordial profiles leads to the "initial average profile". This numerical profile is then taken as an input for the spherical dynamics as studied in \refsec{sec:SC}. We thus obtain a spherically evolved profile which can be compared to the numerical one at $z=0$.

In \refim{fig:spherical_collapse} we plot both this \textit{spherically evolved} profile (from $z\simeq 8$ to $z=0$) and the numerical profile for $R_1=25$ \Mpc. For all redshift, the agreement between the simulation and the spherical evolution is excellent. On small scales however, typically $r\leq 5$ \Mpc, the spherical dynamics fails to almost $5\%$ to $10\%$. It is not surprising that the central over-dense core is not well reproduced by a spherical dynamics but this work focuses on much larger scales where spherical collapse provides an excellent dynamical model. 

\begin{figure*}
\captionsetup[subfigure]{width=0.47\linewidth}
  \begin{center}
    \subfloat[Mass and density contrast at $z=0$ for $R_1=10$ \Mpc. The full lines correspond to the mass contrast $\Delta(r)$ while the dashed curves correspond to $\delta(r)$.]{
      \includegraphics[width=0.5\textwidth]{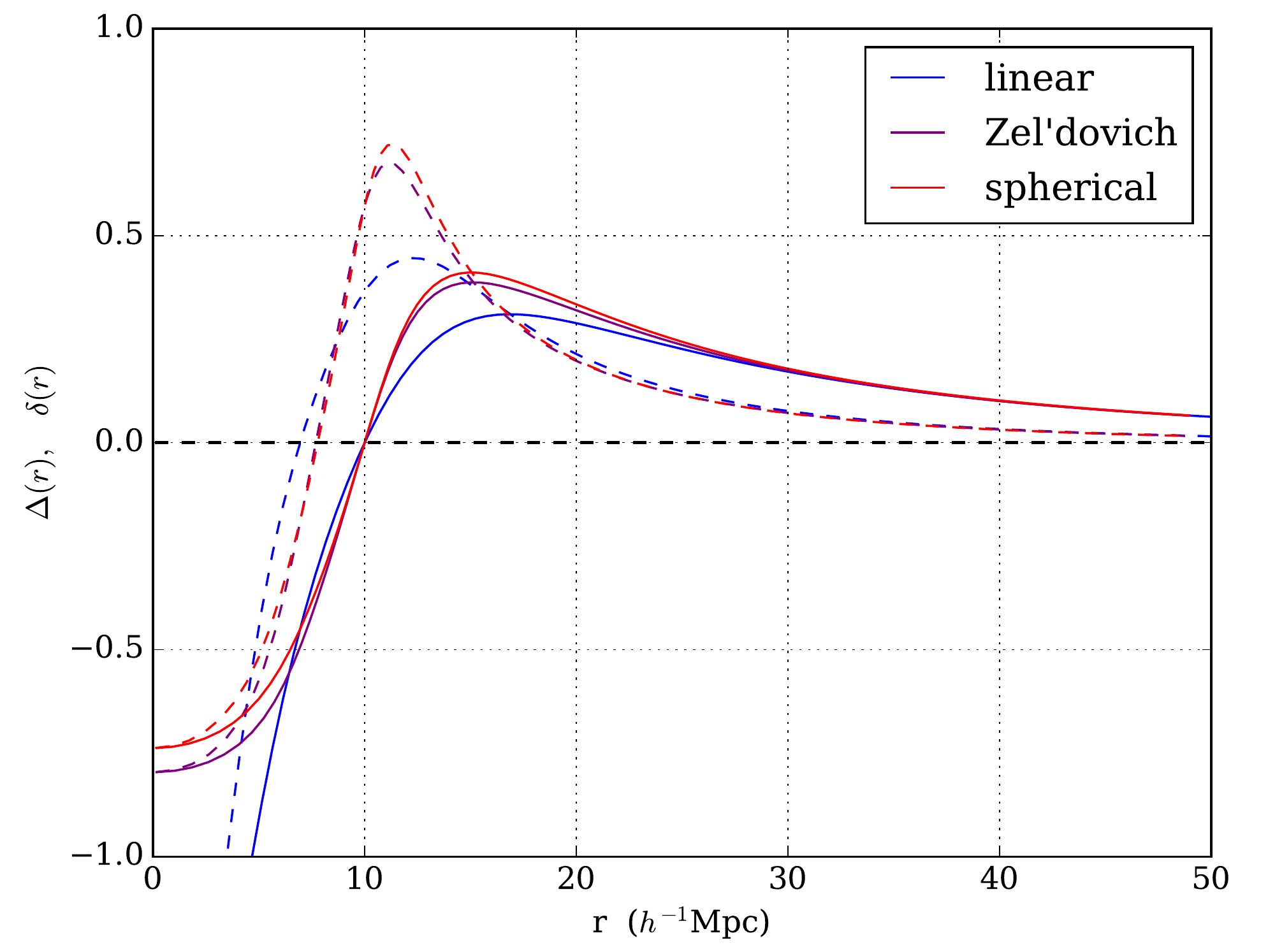}
      \label{sub:mass}}
    \subfloat[velocity contrast profile $\varvel(r)=v/(rH)$ \refeql{varvel}.]{
      \includegraphics[width=0.5\textwidth]{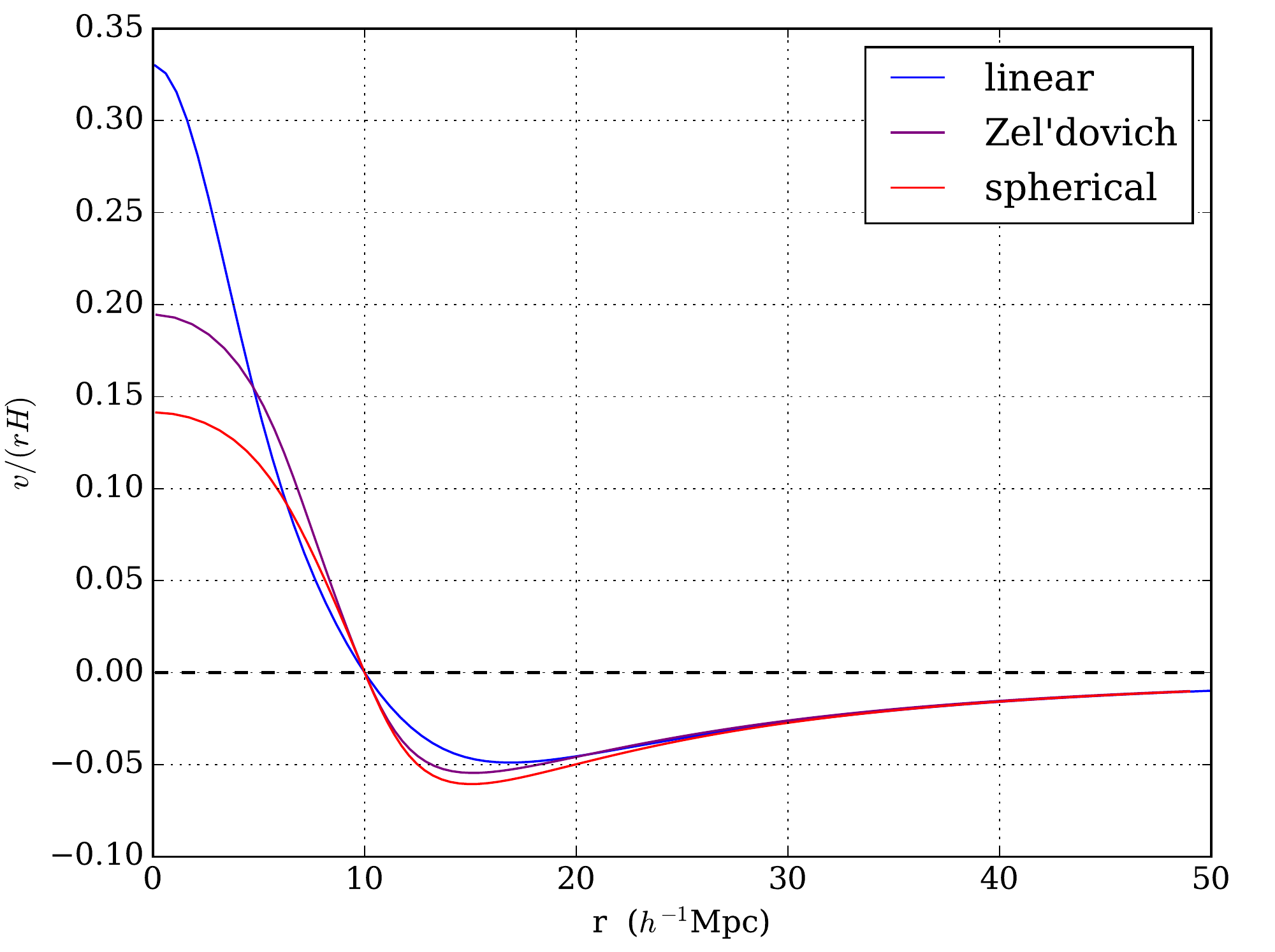}
      \label{sub:velocity}}
    \caption{Comparison of the various dynamical approximations at $z=0$ for a central minimum at $z=0$. Each cure corresponds to the same initial profile evolved with a different dynamical approximation. The Eulerian linear dynamics is in blue, the Zel'dovich approximation is in purple and the full spherical collapse is in red. The deviation appears strongly on velocity profiles than on matter profiles.}
    \label{fig:dynamics_comparison}
  \end{center}
\end{figure*}

On larger scales, such accuracy is neither reachable with the Eulerian linear nor Zel'dovich dynamics. On figure \refim{fig:dynamics_comparison}, we plot the differences at $z=0$ resulting from various dynamical models in the same $\Lambda$CDM cosmology. Here we used a theoretical mass profile computed from our formalism \refsecl{new parametrisation} with realistic shape parameters $\nu$, $x$ and $\nu_1$ (close to unity) and we evolved this profile until $z=0$ for each model. We choose to show these differences on a void profile, \ie central under-dense minima. The argument is exactly symmetric for central over-densities. The Eulerian linear theory (blue lines) is clearly ruled out on non linear scales, \ie for scales typically smaller than $20$ \Mpc. As expected, linear theory, spherical collapse and Zel'dovich approximations agree on linear scales. The Zel'dovich approximation reproduces the spherical dynamics with a precision of $\sim 5\%$ on the mass contrast profile on large scales but only $\sim 10\%$ on the velocity profile. On smaller scales (inside the internal zone, $r<R_1$), the Zel'dovich approximation fails to almost $30\%$. The inaccuracy of the Zel'dovich dynamics cannot be neglected in a precision cosmology era since it could be misinterpreted as a cosmological imprint \citep{paper3,paper4}.

Spherical dynamics is no longer valid in regions where collapse occurred (the shell reaches the singularity $r=0$) and if any shell crosses an other one (\ie when $\partial \Chi/\partial \Chi_i = 0$). But these two limitations are not really relevant for our purpose due to the size of the considered scales. As a matter of fact, the initial radii of shells that collapse in a finite time are very small, typically of the order of the halo size (fraction or order of \Mpc). In the symmetric case of a central under density, the matter field expands such that there is no possible collapse onto $r=0$. Moreover for compensated cosmic regions with realistic values for the shape parameter $\nu$, $x$, and $\nu_1$ (close to unity), radial shell crossing always occurs deep in our future ($z\ll 0$). Note that the shell crossing time $t_{sc}$ of each shell can be easily computed given the initial profile. For example, in the Zel'dovich approximation it satisfies $D(t_{sc})/D(t_i)=1+1/(\delta_i-2/3\Delta_i)$, where both $\delta_i$ and $\Delta_i$ are evaluated at the same initial radius $r_i$.  

\subsection{Dynamical Evolution of the matter field around the compensation radius}

\subsubsection{Evolution of the compensation radius}
\label{sec:R1_evolution}

As was already mentioned in \refim{fig:profile_D_evolution}, the compensation radius is a conserved comoving quantity. This fundamental property is clear from the theoretical point of view. For $r=R_1$ we have $\Delta_i(R_1)=0$ and the only solution of \refeq{dynamic} compatible with the initial conditions \refeq{ini0} and \refeq{ini2} is $\varsc(t)=1$, leading to $R_1\propto a$. 

Physically, since the average density in the closed sphere of radius $R_1$ equals the background density, this sphere evolves exactly as a closed bubble in the Universe and is consequently comoving. 

\begin{figure}
\centering
	\includegraphics[width=1.0\linewidth]{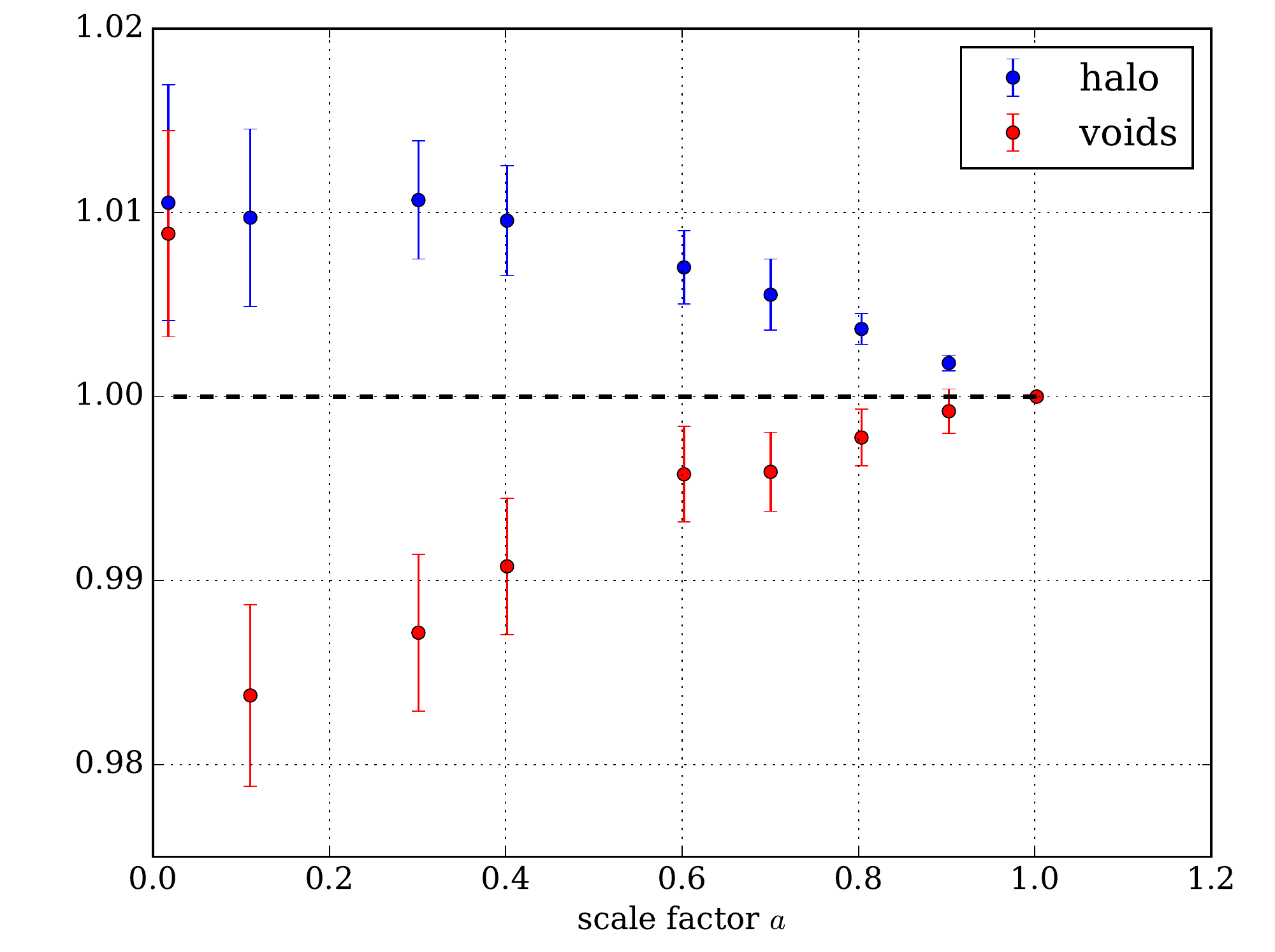}
	\caption{Redshift evolution of the ratio $R_1(z)/R_1(0)$ in comoving coordinates. This figure is obtained using $5000$ profiles around haloes with $M_{h}\sim 1.5\times 10^{13}$ \Msun and the same number of voids without density criteria. For each redshift we compute the distribution for $R_1(z)/R_1(0)$ from which we estimate its mean (points) and the standard error on the mean (error bars). The ratio $R_1(z)/R_1(0)$ is thus conserved with a precision better than $2\%$.}
	\label{fig:r1_z}
\end{figure}

This property stands in a spherical dynamic but initial inhomogeneities are very unlikely to be spherically symmetric \citep{BBKS}. Using numerical simulations, we can follow the redshift evolution of $R_1$ for every CoSphere. In \refim{fig:r1_z} we plot the mean and the dispersion of the distribution $R_1(z)/R_1(0)$ as a function of redshift for both haloes and void. For the whole range of redshift, comoving $R_1$ is constant with a precision better than $2\%$ for voids and $1\%$ for haloes. There is however a clear tendency of increasing $R_1$ for voids and decreasing $R_1$ for haloes. These small evolutions result probably from primordial anisotropies. They remain sufficiently small so that the spherical approximation holds. A deeper understanding of this small evolution goes beyond the scope of this paper.

\subsubsection{The evolution of the compensation density $\delta_1$}
\label{sec:d1_evolution}
The compensation density contrast $\delta_1$ defined as $\delta_1:=\nu_1\sigma_0$ is an Eulerian quantity, being defined at a fixed comoving position $\Chi_1=R_1/a$. To derive its Eulerian dynamics, we consider two points initially located at an infinitesimal distance from the compensation radius $\Chi_i^\pm:=\Chi_1 \pm \epsilon$ where $\epsilon\ll 1$. Since we consider two points in the infinitesimal range $\pm \epsilon$ we have \refeql{zeldo0}
\begin{equation}
\varsc^\pm(t)=1-\left(\pm\epsilon\right)\frac{\delta_1}{\Chi_1}\left(\frac{D(t)}{D(t_i)}-1\right)+\order{\epsilon^2}
\end{equation}
where $\delta_1$ is assumed to be the value of the local density contrast \textit{in the initial conditions}. The same quantity at any time $t\neq t_i$ is explicitly noted with its time dependency $\delta_1(t)$. At first order in $\epsilon$, the position of each shell $\Chi^\pm(t)$ at any time $t$ reduces  to
\begin{equation}
\Chi^\pm(t)=\Chi_1\pm\epsilon\left[1-\delta_1\left(\frac{D(t)}{D(t_i)}-1\right)\right]+\order{\epsilon^2}
\end{equation}
Using the mass conservation \refeq{fnoSC}, the mass contrast for each shell is
\begin{equation}
\Delta^\pm(t) = \pm\epsilon\frac{3\delta_1}{\Chi_1}\frac{D(t)}{D(t_i)}+\order{\epsilon^2}
\end{equation}
Using $\delta_1(t)=R_1/3\Delta'(R_1)$ together with $\Delta'(R_1)=\lim_{\epsilon\to 0}(\Delta^+-\Delta^-)/(\Chi^+-\Chi^-)$, we get
\begin{equation}
\label{deltat}
\delta_1(t)=\delta_1\frac{\tilde{D}(t)}{1-\delta_1\left(\tilde{D}(t)-1\right)}
\end{equation}
With the normalized linear growth factor $\tilde{D}(t)=D(t)/D(t_i)$. This solution corresponds to a one dimensional Zel'dovich dynamics \citep{Zeldovich1970}. However this solution is exact within the spherical collapse. Note that for an initial negative compensation density ($\delta_1<0$ which corresponds to a central over-density, \ie a maximum), the asymptotic value is $-1$ and $\delta_1(t)$ does not diverges to $-\infty$ as expected in the linear regime. The linear regime is recovered for $\delta_1[1-\tilde{D}(t)]\ll 1$ where \refeq{deltat} reduces to the usual linear relation
\begin{equation}
\label{deltalin}
\delta_1(t)\simeq \delta_1 \tilde{D}(t)
\end{equation}
We stress that \refeq{deltat} applies only for the very particular radius $r=R_1$ and cannot be extended to every point of the density profile where it would be, at best, a dynamical approximation. \refeq{deltat} contains its own information about shell-crossing. Indeed, for the particular time $t_{sc}$ satisfying
\begin{equation}
\label{sc_R1}
\frac{D(t_{sc})}{D(t_i)}=\frac{1+\delta_1}{\delta_1}
\end{equation}
The denominator of \refeq{deltat} vanishes, leading to $\delta_1(t_{sc})\to\infty$. This divergence is only possible for positive values of $\delta_1$, \ie for central under-density\footnote{remember that the sign of $\delta_1$ is the opposite of the sign of the central extremum}. The divergence of the local density illustrates the apparition of caustics in the density field, due to the possible crossing of different shells at this particular radius where matter accumulates. 

High values of $\nu_1$ can lead to a collapse of the surrounding over-dense belt onto the central minimum. This is known as the \textit{void-in-cloud} problem \citep{Sheth2004}. For such voids, the compensation belt is deeply affected by shell crossing and the compensation radius is no longer conserved (it decreases with time). The smallest $\nu_1$ leading to radial shell-crossing today depends on $R_g$, the Gaussian smoothing scale of the power spectrum. In $\Lambda$CDM cosmology, this critic $\nu_1$ (computed from \refeq{sc_R1}) evolves from $\nu_1\simeq 0.2$ for $R_g\to 0$ and crosses $\nu_1= 1$ for $R_g\simeq 4$ \Mpc. This illustrates that the shell-crossing mechanism behaves differently according to the smoothing scale. In our case, the power-spectrum is smoothed on the scale equivalent to the size of the coarse grid cell of the simulation to cut the power on smaller scales (in the reference simulation we have $R_{cell}=1.26$ \Mpc). For this smoothing size, the shell-crossing threshold is $\nu_1\simeq 0.65$. As will be shown in \citet{paper2}, this value is much larger than the typical values of $\nu_1$, which are expected to be less than $\nu_1\sim 0.03$. This insures that spherical shell-crossing is very rarely reached in voids and the void-in-cloud effect can thus be neglected, excepted for some very rare events.

Note that this is not in contradiction with the most common definition criterion for voids, namely that they are enclosed by shell-crossed boundaries \citep{Bertschinger1985, Sheth2004}. Indeed, a was already pointed out in \citet{Sheth2004}, this shell-crossing does not appears in sufficiently smoothed profiles, which is the case for realistic CoSphere profiles. The clumpy structuration on small scales where shell crossing locally happened to form virialized structures is not relevant for spherical averaged profiles due to the large volume of radial shells. For central maximum, the shell at $r=R_1$ acts as a gravitational repeller, avoiding caustic formation.

\subsubsection{The local velocity field}
\label{sec:dv1_evolution}
Since the Lagrangian displacement $\varsc$ obeys a second order differential equation \refeql{dynamic}, the field is fully characterized by $\varsc$ and its first derivative. In other word, the radial peculiar velocity (linked to the time derivative of $\varsc$) carries a complementary information. We thus defined the velocity contrast $\varvel$ as
\begin{equation}
\label{varvel}
\dot{r}=rH(t)\left[1 + \varvel(r,t)\right]
\end{equation}
measuring the radial peculiar velocity in units of the Hubble flow $rH$. This dimensionless quantity is computed in the Lagrangian formalism as
\begin{equation}
\label{speed2}
\varvel(r,t):=\frac{\partial\log \varsc}{\partial\log a} = \sqrt{\frac{\Omega_m}{2}}\frac{\partial\log(\varsc)}{\partial \vart}
\end{equation}
and satisfies $\varvel(R_1)=0$ during the whole evolution. In the Zel'dovich regime, mass and velocity contrast profiles are directly proportional
\begin{equation}
\label{speed3}
\varvel(r,t)=-\frac{\Delta(r,t)}{3}\times f(t)
\end{equation}
where $f(t)$ is the linear growth rate. We also define the velocity divergence  $\delta_{vel}(r)=\boldsymbol{\nabla}.\boldsymbol{v}/(3H)$  linked to the velocity contrast by
\begin{equation}
\label{secondary_velocity_contrast}
\varvel'(r)=\frac{3}{r}\Big[\delta_{vel}(r)- \varvel(r)\Big]
\end{equation}
Using a similar computation than in \refsec{sec:d1_evolution}, we can compute the exact non linear evolution of $\delta_{vel}$ around $R_1$
\begin{equation}
\delta_{vel}(R_1,t)=-\frac{f(t)}{3}\frac{\delta_1\tilde{D}(t)}{1-\delta_1\left(\tilde{D}(t)-1\right)}
\end{equation}
With the explicit expression for $\delta_1(t)$ \refeql{deltat}, we get 
\begin{equation}
\label{growth_rate_measure}
\frac{\delta_{vel}(R_1,t)}{\delta_1(t)}=-\frac{f(t)}{3}
\end{equation}
which is the standard relation linking the velocity divergence and the density field in the linear regime. However, in the spherical collapse model, it is an exact result at any redshift for $r=R_1$. For other radii, the previous relation \refeq{growth_rate_measure} is only valid in linear regime.

\refeq{growth_rate_measure} provides an efficient way to evaluate exactly the linear growth rate using CoSpheres. We emphasize that \refeq{growth_rate_measure} allows to measure the linear growth rate on non linear scales, it only necessitate to consider structures will small compensation radii. The measure of the linear growth rate, for example from redshift-space distortions, is beyond the scope of this paper and will be investigated in a forthcoming paper \citep{paper3}.

\subsection{Reconstructing profiles at $z=0$}
\label{sec:reconstruction}

In \refsec{sec:initial} we have shown that the large scale matter profile of CoSpheres can be precisely reconstructed using GRF \refeql{Delta_i}. Theoretical profiles are parametrized by three independent shape parameters $\nu$, $x$, $\nu_1$ \refsecl{new parametrisation} in addition to the compensation radius.

In \refsec{sec:spherical_dynamics} we have shown that the spherical collapse model provides a good description of the gravitational evolution of these large scale profiles. Combining the initial conditions and the dynamics, we show in this section that CoSpheres can be precisely reconstructed until $z=0$ with a high accuracy on a large radial domain.

At $z=0$ we build the average mass contrast profiles of CoSphere in numerical simulations \refsecl{sec:construction}. For each $R_1$ it provides an average profile together with its dispersion (computed as the standard error on the mean). For each profile at $z=0$, the reconstruction procedure consists in finding the appropriate shape parameters $\nu$, $x$ and $\nu_1$ in GRF \refeql{Delta_i}. Practically, we iterate over the shape parameters and minimize a standard $\Chi^2$ at $z=0$ using the spherical evolution of the GRF expected profile.

\begin{figure*}
  \begin{center}
    \subfloat[mass contrast profile for $R_1=20$ \Mpc]
    {
      \includegraphics[width=0.5\textwidth]{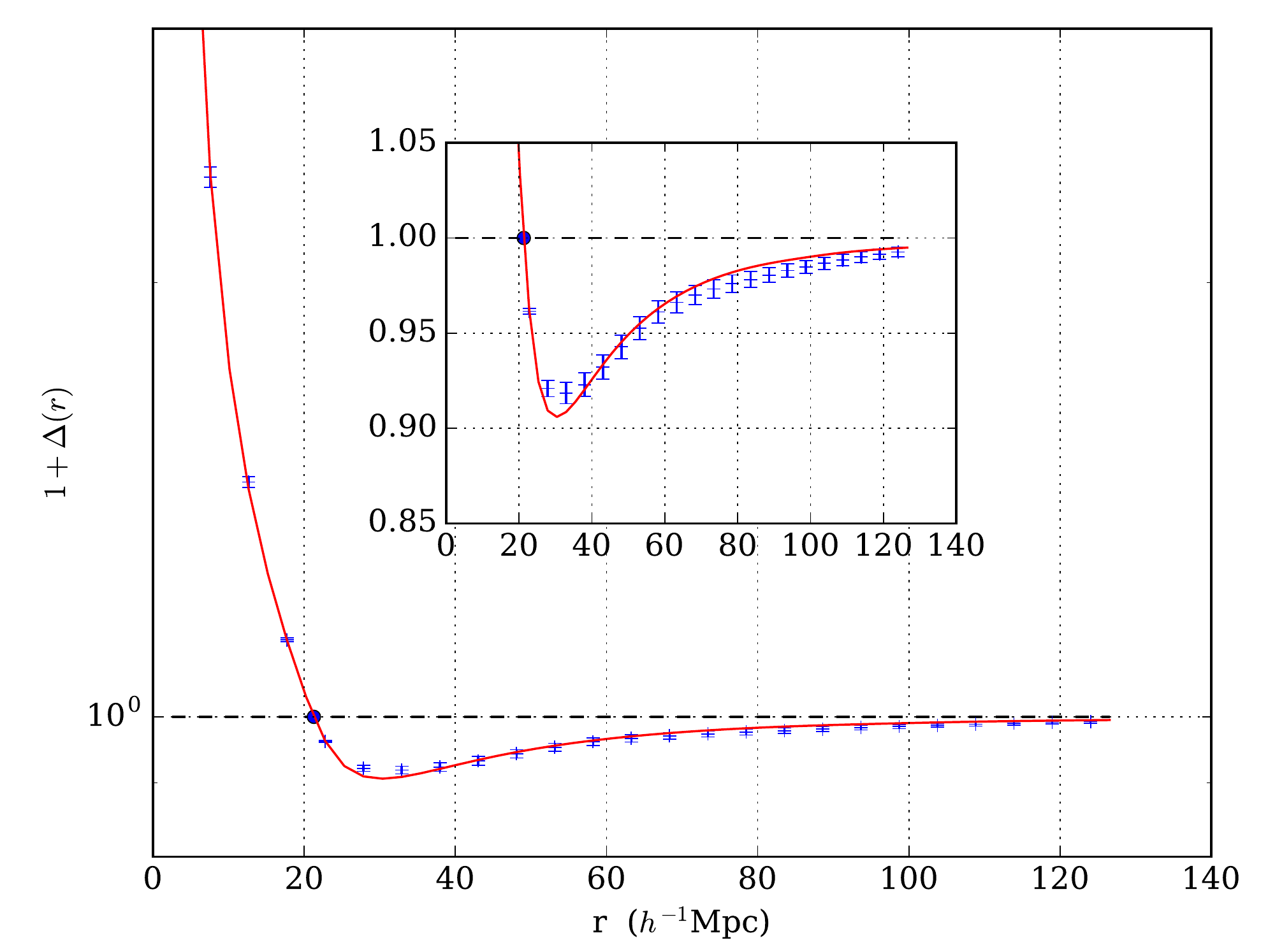}
      }
    \subfloat[same as the left panel for $R_1=40$ \Mpc]
    {
      \includegraphics[width=0.5\textwidth]{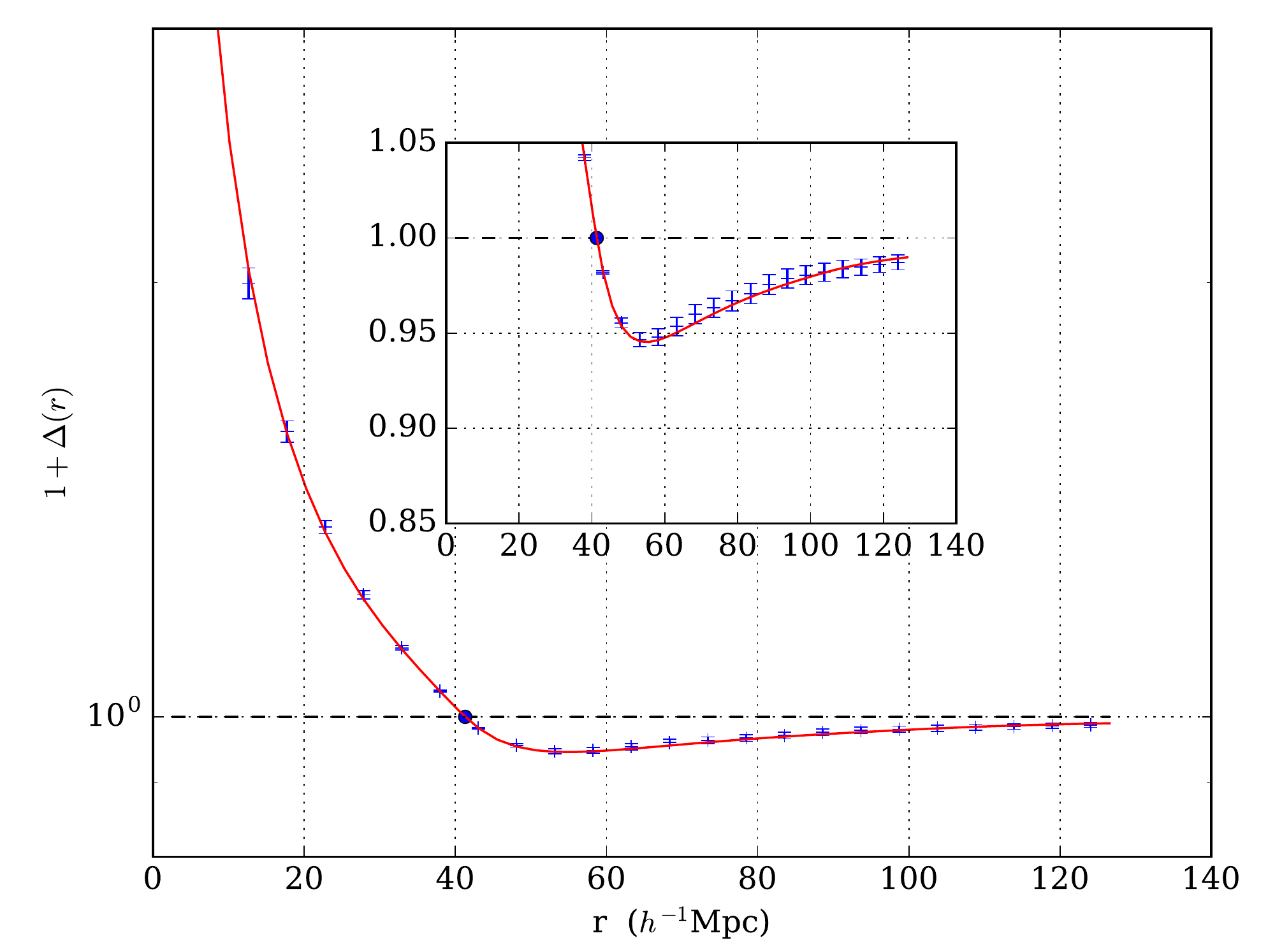}
      }
    \caption{Reconstruction of the mass contrast profile $1+\Delta(r)$ measured in the simulation at $z=0$ (blue points) for haloes of mass $M_h\sim 3.0\times 10^{13}$ \Msun. The red line is the CoSphere curve obtained by minimizing a standard $\Chi^2$ at $z=0$ (see text).}
    \label{fig:fit1}
  \end{center}
\end{figure*}

In \refim{fig:fit1} we show the reconstructed mass contrast profiles at $z=0$ for $R_1=20$ and $R_1=40$ \Mpc. The reconstructed CoSphere reproduces the numerical profile with a very high accuracy (a deviation smaller than $1 \%$) on a large spatial domain. Again we emphasize that the peak parameters $\nu$ and $x$ provide the description of the field around the central extremum whereas $\nu_1$ defined at $r=R_1$ drives the shape on larger scales (see \refim{fig:profile_D_parameters}). \refim{fig:fit1} shows that the reconstruction procedure works for various neighbourhoods. Although we considered the same haloes (same mass), we probe various neighbourhoods by varying the compensation scale $R_1$. A large compensation radius describes a local extremum located in a huge over/under massive region. On the other hand, the same peak with a smaller $R_1$ corresponds to a local extrema in an small over-dense "island" isolated in a larger under-dense region.

\begin{figure*}
	\captionsetup[subfigure]{width=0.47\linewidth}
	\begin{center}
		\subfloat[mass contrast profile for haloes of mass $M_h\sim 3.6\times 10^{12}$ \Msun. We used here the simulation with $1024^3$ particles and a box size $L=648$ \Mpc]{
			\includegraphics[width=0.5\textwidth]{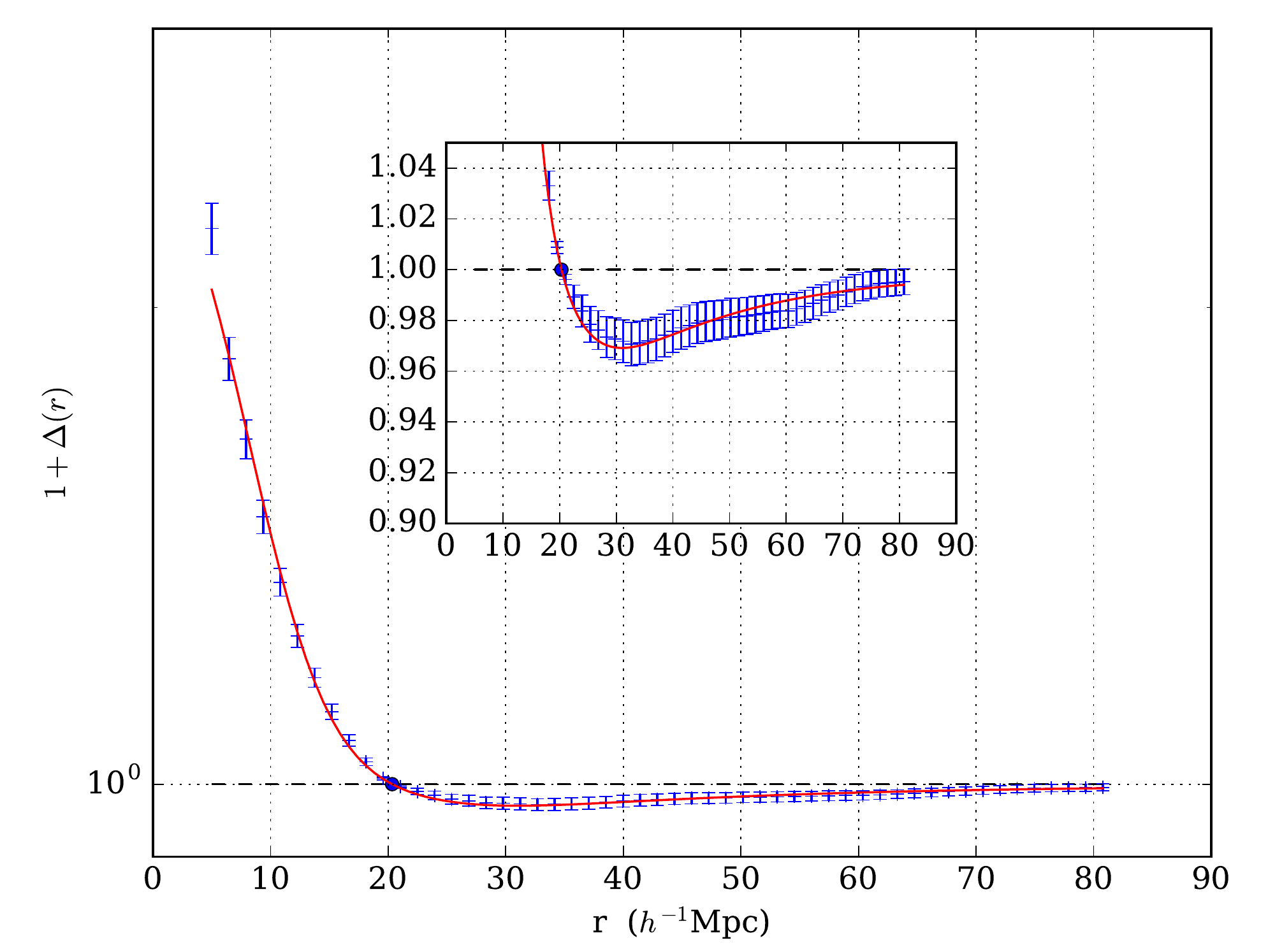}
			\label{sub:small}
		}
		\subfloat[same as the left panel for $M_h\sim 2.5\times 10^{14}$ \Msun with the same $R_1$ in the simulation with $2048^3$ particles and a box size $L=5184$ \Mpc]{
			\includegraphics[width=0.5\textwidth]{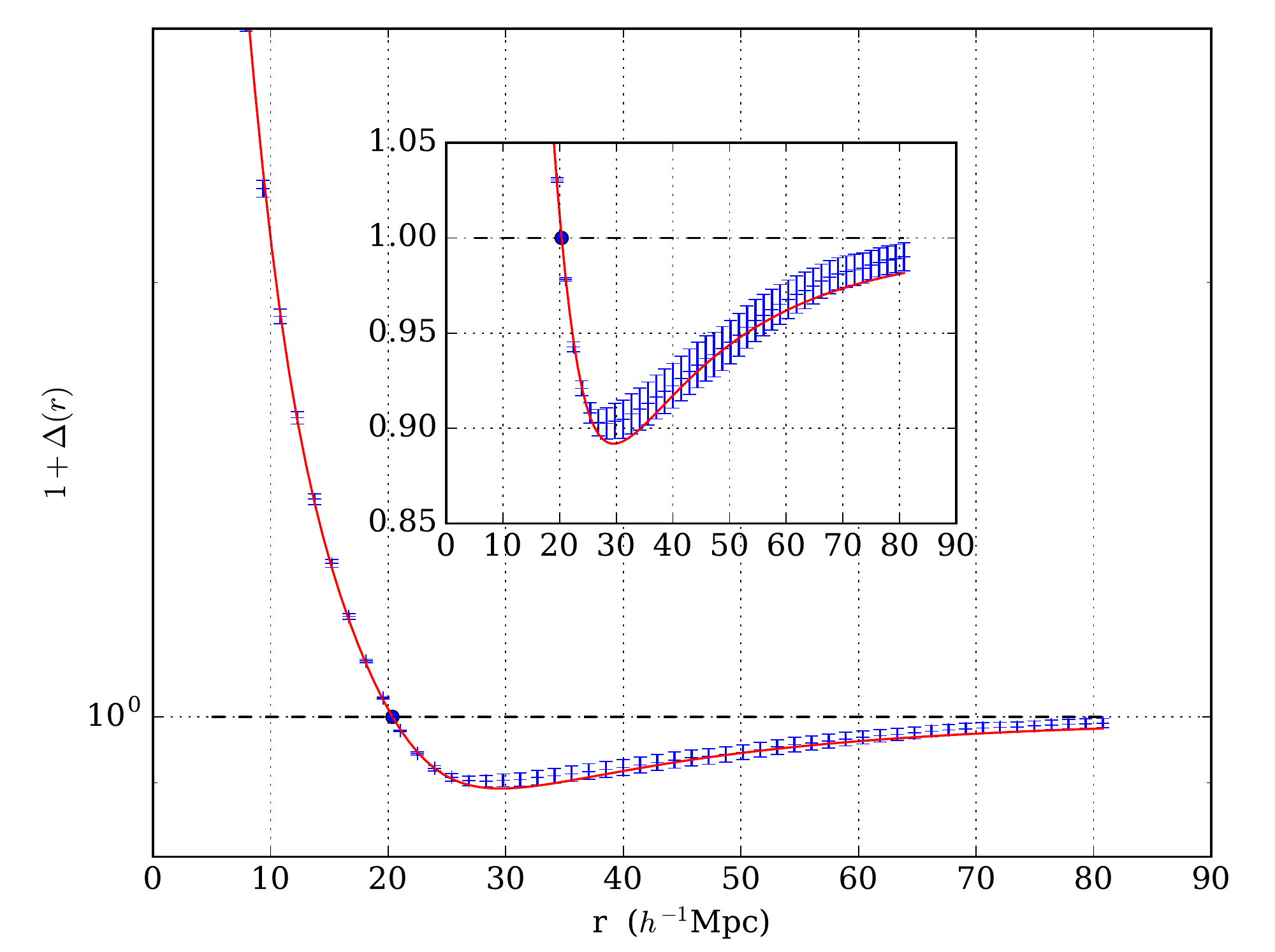}
			\label{sub:high}
		}
		\caption{Reconstruction of CoSphere profile at $z=0$ (blue points) from two different haloes with the same compensation radius $R_1=20$ \Mpc. The red line is the theoretical curve obtained by computing the best shape parameters $\nu$, $x$ and $\nu_1$ and spherically evolved until $z=0$ (see text). This figures illustrate the mass dependence of the profiles. Matter profiles around heavier haloes are more amplified than the same profiles build from lighter ones. Increasing the mass of the central haloes raises the primordial height $\nu=\delta(\bx_0)/\sigma_0$. Shape parameters are correlated to each other such that it is more likely to get higher $\nu_1$ when $\nu$ growths \citep{paper2}. Heavier haloes will thus induce more amplified profiles on all scales, as it is illustrated in this figure.}
		\label{fig:fit2}
	\end{center}
\end{figure*}

On \refim{fig:fit2} we show the reconstruction of CoSphere profiles defined around haloes with different masses, namely $M_h\sim 3.6\times 10^{12}$ \Msun (extracted from the simulation with $1024^3$ particles and a box size $648$ \Mpc) and  $M_h\sim 2.5\times 10^{14}$ \Msun (extracted from the simulation with $2048^3$ particles and a box size $5184$ \Mpc) for the same compensation radius $R_1=20$ \Mpc. Varying the mass of the central halo changes the amplitude of matter fluctuation, and thus the profile itself. Increasing the mass of the central halo raises the primordial peak threshold, \ie selects peaks with higher $\nu$. Since the central extrema is correlated to its surrounding environment, large $\nu$ induce higher $\delta_1$ and thus $\nu_1$. In other words, a massive halo is more likely to sit in a deepest void than a lighter halo. 

\begin{figure*}
  \begin{center}
    \subfloat[mass contrast profile for $R_1=20$ \Mpc]{
      \includegraphics[width=0.5\textwidth]{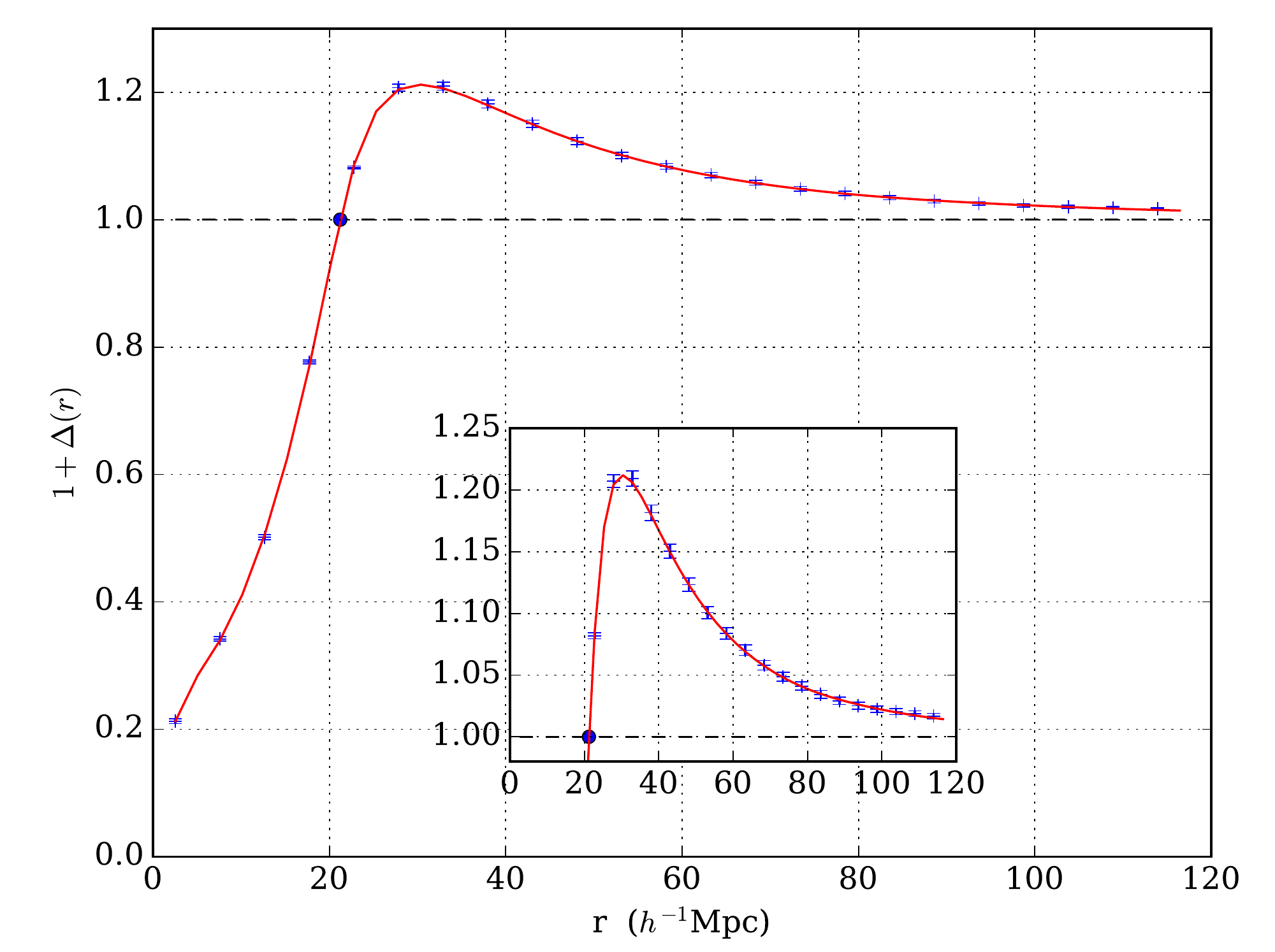}
      \label{sub:full}}
    \subfloat[same as the left panel for $R_1=40$ \Mpc]{
      \includegraphics[width=0.5\textwidth]{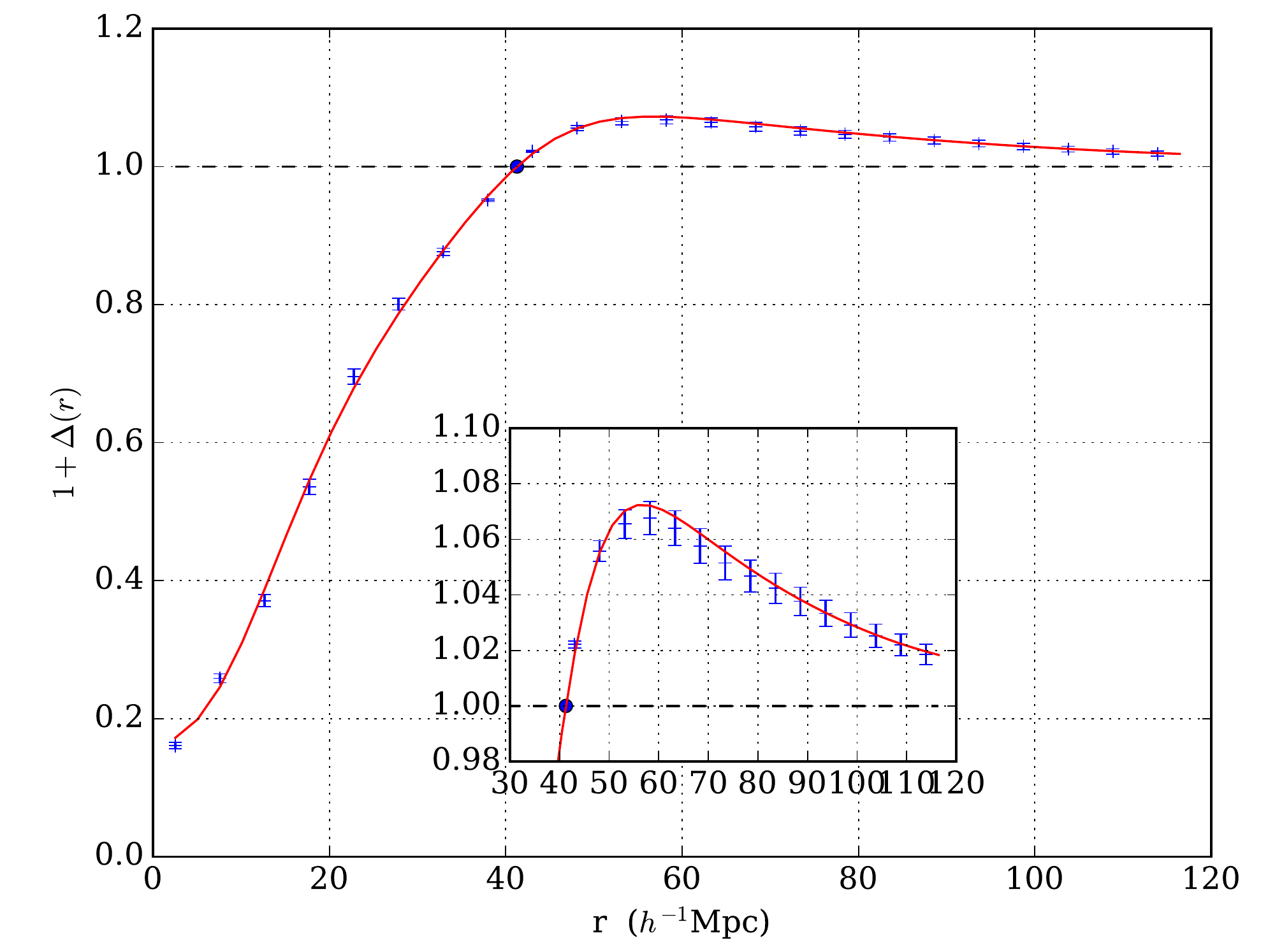}
      \label{sub:zoom}}
    \caption{Reconstructed CoSphere profile at $z=0$ around local under-dense minimum. These minima are obtained by smoothing the density field with a Gaussian kernel with $R_g=2$ \Mpc. The red line is the theoretical curve obtained from GRF with the best fit shape parameters and evolved with a spherical dynamics (see text). It is noticeable that the reconstruction provides an excellent fit on all scales and whatever $R_1$ although theoretical profiles are determined by three parameters including two parameters defined around $r=0$.}
     \label{fig:fit_void}
  \end{center}
\end{figure*}

Finally, the exact same reconstruction can be done for central under-dense regions identified to cosmic voids. The measured profiles of under-dense CoSphere together with their theoretical reconstruction are shown in \refim{fig:fit_void}. We show the reconstruction for two different compensation radii. Here again, CoSphere profiles are well reconstructed on all scales with a very high accuracy, even in the central under-dense core ($r\ll R_1/5$). This is not surprising since cosmic voids tends to sphericity during the tri-axial expansion, unlike their over dense symmetric \citep{Icke1984, Weygaert2014}. CoSpheres provide thus an efficient physically motivated model that can be used to reproduce large scale spherical inhomogeneities at any redshift. 

\section{Discussion and outlooks}

The absence of a physically motivated model for understanding the large scale matter profiles of compensated cosmic regions is a major difficulty in the precision cosmology era. Extracting reliable cosmological information from such regions, and particularly from voids, requires a deep understanding of their origin and their evolution. In this paper we address this issue by generalizing void profiles and introducing CoSpheres. These regions are build explicitly from their compensation property. The particular radius $R_1$ where the matter field compensate exactly appears to be a fundamental scale for both their origin and their dynamics. This comoving radius isolates closed bubble Universe with a conserved volume during the whole cosmic evolution \refsecl{sec:R1_evolution}. 

When defined around central under dense minimum, these regions can be identified to cosmic void, providing a useful theoretical framework for studying both their shape and their evolution. Interestingly, these regions can be also defined around local maximum such as DM haloes. By definition, these regions must be compensated on a finite scale, hence the existence of large under dense regions surrounding over densities. 

Using numerical simulations introduced in \refsec{sec:DEUS} we build the averaged profiles of CoSpheres by stacking together regions with the same compensation radius $R_1$. These numerical simulations can be used to follow backward in time the evolution of such cosmic structures \refsecl{sec:initial_reconstruction}. From these primordial numerical profiles we have shown that CoSpheres are generated from the stochastic fluctuations of the primordial field (see \refsec{sec:cosphere_initial} and \refim{fig:profile_D_evolution}).

At high redshift the matter field follows a Gaussian statistics. In order to derive the matter profile of CoSpheres in GRF formalism, we have extended the results of \citet{BBKS} by implementing explicitly the compensation conditions \refeq{R1} and \refeq{v1} \refsecl{sec:initial}. With this original compensation constraint, the spherical density (and mass) contrast profile is now parametrized by four independent - but correlated - shape parameters; $\nu$ and $x$ qualifying the central extrema (already introduced by BBKS) while $\nu_1$ and $R_1$ characterize the surrounding cosmic environment on larger scales. While the standard BBKS profile was determined on all scales by providing the peak parameters $\nu$ and $x$, our extension allows to probe the same central extremum in various cosmic environments. These physical configurations can be described by the additional shape parameters $\nu_1$ and $R_1$ (see \refim{fig:profile_D_bbks_comparison}). We emphasize that $\nu$ and $x$ affect the matter profile on small scales while $\nu_1$ controls the shape and the amplitude on larger scale, typically around and beyond the compensation radius. 

In \refsec{sec:spherical_dynamics} we show that the spherical collapse model is well suited for the dynamical evolution of CoSpheres whereas neither Zel'dovich nor linear dynamics provide satisfying accuracy. We show that the full non linear gravitational collapse can be solved analytically around $R_1$ where it reduces to a one dimensional Zel'dovich dynamics. We stress that on this particular radius, the Zel'dovich dynamics provides an exact solution for the spherical collapse and not only a dynamical approximation. In particular, this implies that the linear growth rate can be exactly estimated on this scale \refeql{growth_rate_measure}. This emphasize the relevance of this particular radius. The possibilities to use this property and to constrain the underlying cosmology will be discussed in detail in \citet{paper3, paper4}. 

\begin{figure}
	\includegraphics[width=1.0\linewidth]{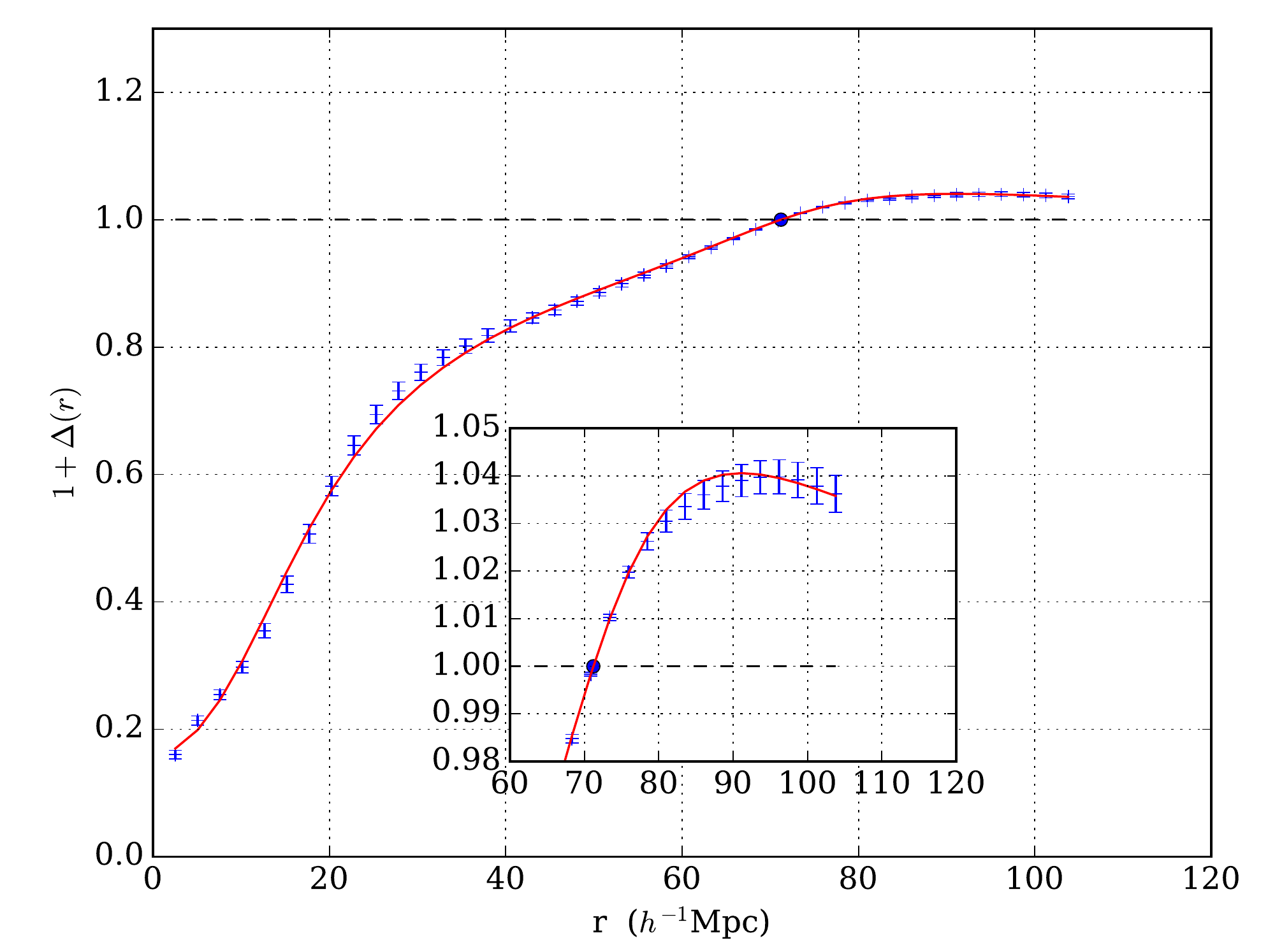}
	\caption{Reconstruction of a void profile at $z=0$ with a large compensation radius $R_1=70$ \Mpc. The red curve is the CoSphere reconstruction obtained through the procedure discussed in \refsec{sec:reconstruction}. This figure illustrates the particular elbow around $r\sim 25$ \Mpc as in \refim{fig:averaged_profile}. This original feature will be discussed in a following paper \citet{paper2} and appears as an imprint of the decoupling between the central extrema and its surrounding environment.}  
	\label{fig:fit_elbow}
\end{figure}

For central minimum, CoSphere can be identified to cosmic voids. Their radial profiles exhibit a characteristic elbow around $r\sim 20$ \Mpc (see \refim{fig:profile_D_void_today} and \refim{fig:fit_elbow}). This elbow is present on both density and mass contrast profile, though it is more pronounced on density profiles (red curve on \refim{fig:profile_D_d_today_void}). This particular shape property is a characteristic of the definition of our cosmic voids and does not appear clearly in void profiles build from other algorithm (e.g. \citet{Hamaus2014}). 

This elbow is a specific feature of our stacking operation which combine profiles with the same compensation radius $R_1$. For other void reconstructions based on their effective radius $R_{eff}$, this elbow may be smoothed by the stacking together profiles with various $R_1$. As will be discussed in \citet{paper2}, this elbow is the imprint of the progressive decorrelation between the central extrema and the surrounding cosmic environment.

We stress that our work allows a common description for the formation of both cosmic void and large scale profile surrounding haloes. The efficiency of the reconstruction procedure \refsecl{sec:reconstruction} emphasizes $R_1$ as a fundamental scale carrying the memory of the primordial Universe and qualifying cosmic structures.

\begin{figure}
	\includegraphics[width=1.0\linewidth]{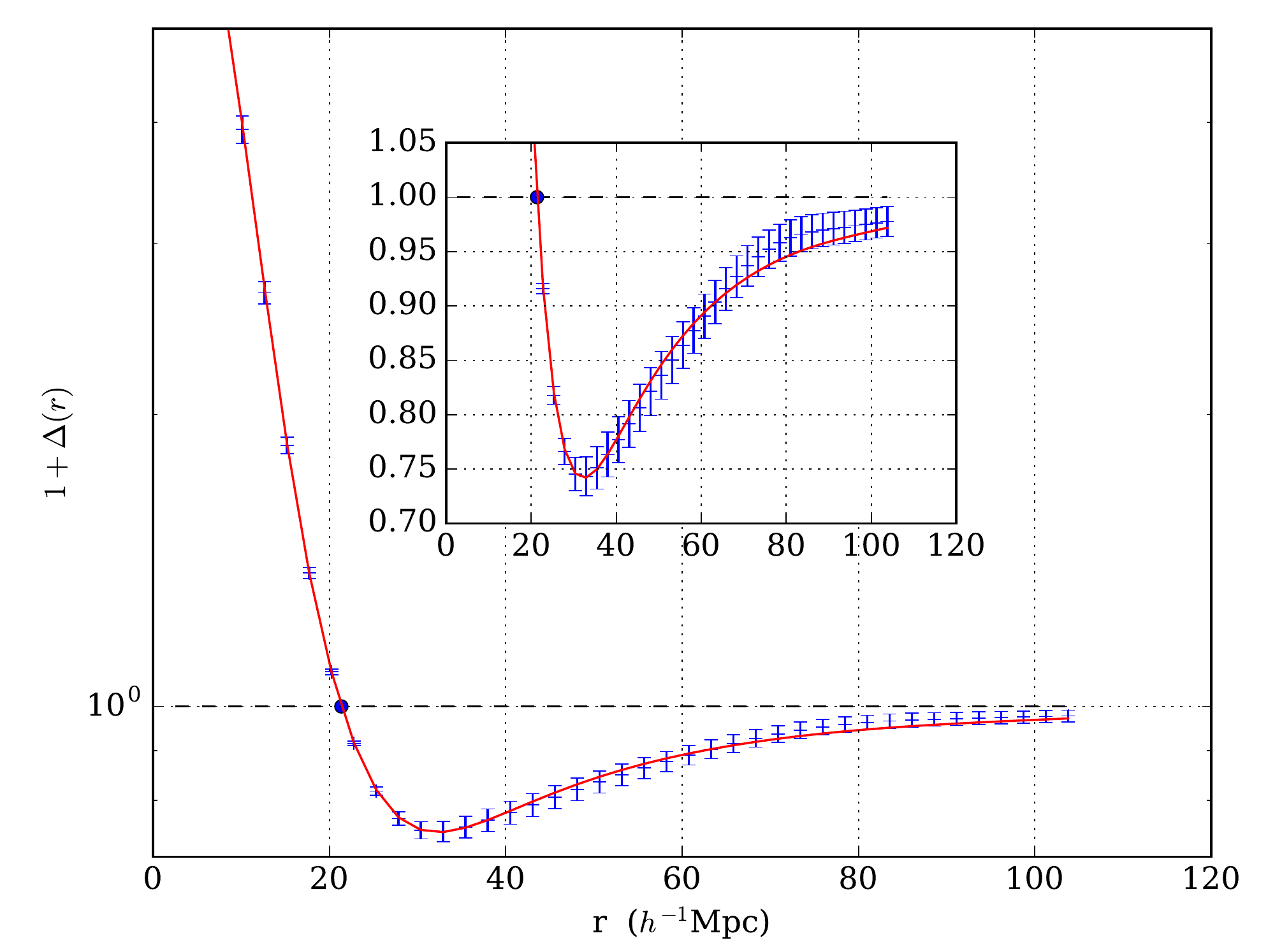}
	\caption{Mass contrast profile from central haloes of mass $M_h\sim 3.0\times 10^{13}$ \Msun computed from the \textit{halo field} at $z=0$. This profile is obtained by considering a biased field traced by DM haloes weighted by their mass. The mean density $\bar{\rho}$ is estimated from the ratio between the total mass of haloes and the volume of the simulations. The red cure is the theoretical reconstruction using CoSphere formalism as discussed in \refsec{sec:reconstruction}}  
	\label{fig:fit_halo}
\end{figure}

Finally, all the results presented in this paper assume a non biased or distorted CDM field. In realistic surveys however, we don't have access to the full CDM field but rather to its discrete tracers as galaxies or galaxy cluster. In a N-body simulation, these tracers can be modelled from Dark Matter haloes since galaxies are more likely to form in potential wells generated by DM collapse. As a proof of concept we show in \refim{fig:fit_halo} the reconstructed matter profile obtained from the field traced only by DM haloes. The global shape of profiles is not changed when using the biased field and CoSpheres can still be clearly identified. The agreement between numerical profiles (blue points) and reconstructed theoretical profile (in red) is again excellent on all scales. The only modification with the previous matter profiles reduces, in a first approximation, to the introduction of a linear bias $b$ such as $\delta_{haloes}=b\times \delta_{CDM}$ without affecting the shape of CoSpheres. For cosmic voids, the linearity of the bias has been studied in \citet{Pollina2017} where it was shown to be a very good approximation, whatever the tracer population (galaxies, galaxy clusters and AGN).

\section*{Acknowledgements}
We thank the anonymous referee for useful comments. We also thank Jean Pasdeloup for his help in developing numerical tools for void detection in numerical simulations.

\bibliographystyle{mnras}
\bibliography{biblio}

\appendix

\section{Shifting the center of mass}
\label{shift}
The primordial profiles computed in \refeq{Delta_i} and \refeq{profile_0} have been derived within the GRF formalism assuming that the position $\bold{x}_0$ of the central extrema is perfectly known. However, the reconstruction procedure of the averaged profiles at higher redshift as introduced in \refsec{sec:initial_reconstruction} may induce a shift between the exact position of the local extrema $\boldsymbol{x}_0$ and its estimation $\boldsymbol{x}_c$. In this section we evaluate the consequence of missing this exact position. 

We assume that the density profile centred on the real extrema $\avg{\delta}(r_i)$ is given by \refeq{profile_0} where $r_i$ denotes the comoving distance from the real extrema $\bold{x}_0$, \ie $r_i=\abs{\bold{x}_0-\bold{x}}$. We want to evaluate the density contrast on the shell located at a radius $r$ where $r$ is measured from the estimated (but \textit{'wrong')} center $\bold{x}_c$ shifted by $\bold{x}_c:=\bold{x}_0+\bold{R}$. In other words we note $r=\abs{\bold{r}}=\abs{\bold{x}_c-\bold{x}}$. 

Let us define $\varphi$ such as $\bold{R}.\bold{r}=R.r.\cos(\varphi)$. Following \refeq{profile_0}, the spherical density contrast can be written as an Hankel Transform
\begin{equation}
\label{hankel}
\avg{\delta}(r_i)=\int P(k)\times \tilde{\delta}(\nu,x,\nu_1,R_1,k)\frac{\sin(kr_i)}{kr_i}dk
\end{equation}
Where $\tilde{\delta}$ is a linear function of $\nu$, $x$ and $\nu_1$ and depends non linearly in $k$ and $R_1$ \refeql{profile_0} while $P(k)$ is the linear power-spectrum. The "reconstructed" density contrast $\avg{\delta}'(r)$ around the position $\bold{x}_c$ at a radius $r$ is thus given by averaging $\avg{\delta}$ on the shifted sphere of radius $r$ around $\bold{x}_c$ 
\begin{equation}
\avg{\delta}'(r)=\frac{1}{2}\int_0^\pi\sin(\varphi)\avg{\delta}\left(\sqrt{r^2 + R^2- 2rR\cos(\varphi)}\right)d\varphi
\end{equation}
Using the explicit expression \refeq{hankel} for $\avg{\delta}$ we find 
\begin{equation}
\avg{\delta}'(r)=\int P(k)\times\tilde{\delta}(\nu,x,\nu_1,R_1,k)\frac{\sin(kr)}{kr}\times\frac{\sin(kR)}{kR}dk
\end{equation}
Thus, the shifted profile $\avg{\delta}'(r)$ takes exact same form than the un-shifted profile $\avg{\delta}(r)$ given by \refeq{profile_0} with an effective power-spectrum $P_{eff}(k)$ given by
\begin{equation}
\label{effective_P}
P(k)\underset{\bold{x}_0\to\bold{x}_0 +\bold{R}}{\to} P_{eff}(k):=P(k)\times\frac{\sin(kR)}{kR}
\end{equation}
Of course, for $R\to 0$ we recover the usual profile but for $R\neq 0$, the profile implies an effective power spectrum smoothed on the shifting scale $R$. Note also that missing the right center of mass leads to non-isotropic profiles but here we focus only the the spherically average profile. The mass contrast profile $\avg{\Delta}$ \refeql{Delta_i} is affected by the exact same factor, \ie it is written exactly as \refeq{Delta_i} but with the effective spectrum given by \refeq{effective_P}.

\section{High redshift solution and the Zel'dovich approximation}
\label{sec:Zeldo}

The dynamical equation \refeq{dynamic} can be solved exactly orders by orders for $\varsc$ (and for each radius $r_i$) with the series
\begin{equation}
\label{expand_solution}
\varsc(t)=1+\sum_{n\geq 1}\eta_n(t)\Delta_i^n
\end{equation}
Where each function $\eta_n(t)$ depends only on $t$ and $\eta_n(t_i)=0$. The solution \refeq{expand_solution} is the exact solution for the Lagrangian perturbation theory in spherical coordinates which is valid until shell-crossing.

\subsection{High redshift solution}
In the very high redshift regime ($z\gg 1$), the initial mass contrast $\Delta_i$ satisfies $\Delta_i\ll 1$ for all initial radius $r_i$ (since $\Delta_i\sim \sigma_0$). The $0$-th order term of \refeq{expand_solution} corresponds to the linear Eulerian theory $\delta(\boldsymbol{x}, t)\propto D(t)\delta(\boldsymbol{x}, t_i)$.

Let us now  consider the first order term
\begin{equation}
\label{zeldo0}
\varsc(t)\simeq 1 + \eta_1(t) \Delta_i + \mathcal{O}(\Delta_i^2)
\end{equation}
with $\eta_1(t_i)=0$. In this regime, the right-hand term of \refeq{dynamic} reduce to
\begin{equation}
\label{solution1}
\varsc-\frac{1+\Delta_i}{\varsc^2}\underset{\Delta_i\ll 1}{\to}\Delta_i\Big(3\eta_1(t) - 1\Big)
\end{equation}
If we define $J$ such as $\eta_1  = (J-1)/3$, using \refeq{solution1}, it is easy to show that $J(t)$ satisfies
\begin{equation}
\label{wfr1}
\frac{d^2J}{dt^2}+2H(t)\frac{dJ}{dt}=\frac{3}{2}H^2(t)\Omega_m(t)J(t)
\end{equation}
With $J(t_i)=1$. \refeq{wfr1} is exactly the equation solved by the linear growth factor $D(t)$, thus, using $J(t_i)=1$ we deduce that in this weak field regime, $\eta(t)$ is given by
\begin{equation}
\label{wfr2}
\eta_1(t)=-\frac{1}{3}\left(\frac{D(t)}{D(t_i)}-1\right)
\end{equation}
In this regime, the displacement field $\varsc$ is given by
\begin{equation}
\label{wfr3}
\varsc(t)=1-\frac{\Delta_i}{3}\left(\frac{D(t)}{D(t_i)}-1\right)
\end{equation}
Where the $\Delta_i$ and thus $\varsc$ depend on the initial position $r_i$.
\subsection{Link with the Zel'dovich approximation}
\label{Appendix_Zeldo}
The Zel'dovich approximation (denoted as ZA) \citep{Zeldovich1970} consists into approximating the field displacement by its initial value. With our notation, it reaches
\begin{equation}
\label{ZA1}
\bold{\Chi}(\bold{q},t)=\bold{q}+\bold{s}(\bold{q},t)
\end{equation}
where $\bold{s}(\bold{q},t)$ is the displacement field which verifies
\begin{equation}
\label{ZA3}
\frac{\partial^2\bold{s}}{\partial t^2}+2H\frac{\partial \bold{s}}{\partial t}=-\bold{\nabla}\phi
\end{equation}
and $\phi$ is the gravitational potential that satisfies $\Delta\phi=4\pi G\delta(r)$. The ZA approximates the displacement field by its initial value $\bold{s}(\Chi_i,t)=\bold{s}_0(\bold{q})D(t)$ where $D(t)$ is the linear growth factor which verifies \refeq{wfr1} and $\bold{s}_0 = -(2\bold{\nabla}\phi_0)/(3H_i^2\Omega_{m,i})$. 

If we define $\Chi_i$ such as $\Chi(\bold{q},t_i)=\Chi_i$, then $\Chi(\Chi_i,t)=\Chi_i+\bold{s}(\bold{q},t)-\bold{s}_0(\bold{q})$. Using \refeq{poisson2}, we can write explicitly $\bold{s}_0(\bold{q})$ in the spherical approximation. It then follows that 
\begin{equation}
\label{ZA4}
\Chi=\Chi_i\left(1-\frac{\Delta_i}{3}\left[\frac{D(t)}{D(t_i)}-1\right]\right)
\end{equation}
which is exactly the solution obtained in \refeq{wfr3}. This is not surprising since the ZA is by construction the first order Lagrangian perturbation theory, we re-find it here in spherical geometry. 
\end{document}